\documentclass[preprint,aps,prd,showpacs,nofootinbib,floatfix,tightenlines]{revtex4}
\usepackage{graphicx}
\usepackage{amsmath}
\usepackage{amssymb}
\usepackage{bm}

\begin{document}

\title{\mbox{}\\[10pt]
The Effects of Charged Charm Mesons \\
on the Line Shapes of the $\bm{X(3872)}$}

\author{Eric Braaten and Meng Lu}
\affiliation{Physics Department, Ohio State University, Columbus,
Ohio 43210, USA}

\date{\today}
\begin{abstract}
The quantum numbers $J^{PC} = 1^{++}$ of the $X(3872)$ 
and the proximity of its mass to the $D^{*0} \bar D^0$ threshold
imply that it is either a loosely-bound hadronic molecule 
whose constituents are a superposition of $D^{*0} \bar D^0$
and $D^0 \bar D^{*0}$ or it is a virtual state of charm mesons.
The line shapes of the $X(3872)$ can discriminate between 
these two possibilities.
At energies within a few MeV of the $D^{*0} \bar D^0$ threshold,
the lines shapes of the $X$ produced in $B \to K$ transitions
are determined by its binding energy and its width.
Their normalizations are determined by a short-distance constant 
that is different for $B^+ \to K^+$ and $B^0 \to K^0$.  
At energies comparable to the 8 MeV splitting between the 
$D^{*0} \bar D^0$ and $D^{*+} D^-$ thresholds,
the charged meson channels $D^{*+} D^-$ and $D^+ D^{*-}$
have a significant effect on the line shapes of the $X$.
We calculate the line shapes taking into account the resonant 
coupling between the charged and neutral $1^{++}$ channels.
The line shapes and their normalizations depend on one additional 
scattering parameter and two additional short-distance constants
associated with the $B \to K$ transitions.  The line shapes 
of the $X$ resonance depend on its decay channel; 
they are different for $J/\psi \, \pi^+ \pi^-$, 
$J/\psi \, \pi^+ \pi^-\pi^0$, and $D^0 \bar D^0 \pi^0$.
The line shapes are also different for $X$ produced in $B^+$ decays 
and in $B^0$ decays.  Some conceptual errors in previous work 
on this problem are pointed out.
\end{abstract}



\maketitle


\section{Introduction}

The $X(3872)$ is a $c \bar c$ resonance near 3872 MeV discovered in 2003
by the Belle Collaboration \cite{Choi:2003ue}
and subsequently observed by the CDF, Babar, and D0 Collaborations
\cite{Acosta:2003zx,Abazov:2004kp,Aubert:2004ns}.
In addition to the discovery decay mode $J/\psi \, \pi^+ \pi^-$, 
the $X$ has been observed to decay into $J/\psi \, \gamma$,
$J/\psi \, \pi^+ \pi^-\pi^0$, \cite{Abe:2005ix} and 
$D^0 \bar D^0 \pi^0$ \cite{Gokhroo:2006bt,Babar:2007rv}. 
The decay into $J/\psi \, \gamma$ implies that the $X$
is even under charge conjugation.  An analysis by the Belle Collaboration
of the decays of $X$ into $J/\psi \, \pi^+ \pi^-$ 
strongly favors the quantum numbers $J^{PC} = 1^{++}$,
but does not exclude $2^{++}$ \cite{Abe:2005iy}.
An analysis by the CDF Collaboration of the decays of 
$X$ into $J/\psi \, \pi^+ \pi^-$ is 
compatible with the Belle constraints \cite{Abulencia:2005zc}.
The tiny phase space available for the decay into $D^0 \bar D^0 \pi^0$
rules out $J=2$, leaving $1^{++}$ as the only option.

An important feature of the $X(3872)$ is that its mass $M_X$ is 
extremely close to the $D^{*0} \bar D^0$ threshold.  
The PDG value for $M_X$ comes from combining 
measurements of $X$ in the $J/\psi \, \pi^+ \pi^-$ decay mode
\cite{Yao:2006px}.  After taking
into account a recent precision measurement of the $D^0$ mass by the
CLEO Collaboration \cite{Cawlfield:2007dw}, the difference between
the PDG value for $M_X$ and the $D^{*0} \bar D^0$
threshold is 
\begin{equation}
M_X - (M_{*0} + M_0) = -0.6 \pm 0.6~{\rm MeV} , 
\label{MX-CLEO}
\end{equation}
where $M_{*0}$ and $M_0$ are the masses of $D^{*0}$ and $D^0$.
The negative central value in Eq.~(\ref{MX-CLEO}) is compatible 
with the $X$ being a bound state of the charm mesons.
The measured mass of the near-threshold enhancement in 
$D^0 \bar D^0 \pi^0$ is about 4 MeV above the $D^{*0} \bar D^0$
threshold \cite{Gokhroo:2006bt,Babar:2007rv}.
This value is compatible with $X$ being a virtual state of charm mesons.
It differs from the mass in Eq.~(\ref{MX-CLEO}) by more than two
standard deviations, which raises the question of whether 
the decays into $J/\psi \, \pi^+ \pi^-$ and  
$D^0 \bar D^0 \pi^0$ are coming from the same resonance.

The proximity of the mass of the $X(3872)$ to the
$D^{*0} \bar D^0$ threshold has motivated its identification as a
weakly-bound molecule whose constituents are a superposition
of the charm meson pairs $D^{*0} \bar D^0$ and $D^0 \bar{D}^{*0}$
\cite{Tornqvist:2004qy,Close:2003sg,Pakvasa:2003ea,Voloshin:2003nt}.
The establishment of the quantum numbers of the $X(3872)$
as $1^{++}$ makes this conclusion almost unavoidable.  
The reason is that these
quantum numbers allow S-wave couplings of the $X$ to
$D^{*0} \bar D^0$ and $D^0 \bar{D}^{*0}$.  
Nonrelativistic quantum mechanics implies that a resonance 
in an S-wave channel near a 2-particle threshold has special 
universal features \cite{Braaten:2004rn}.  Because of the small 
energy gap between the resonance and the 2-particle threshold, 
there is a strong coupling between the resonance 
and the two particles.  This strong coupling 
generates dynamically a large length scale that can be 
identified with the absolute value of
the S-wave scattering length $a$ of the two particles.
Independent of the original mechanism for the resonance, 
the strong coupling transforms the resonance into a 
bound state just below the two-particle threshold if $a>0$ 
or into a virtual state just above the two-particle threshold if $a<0$.
If $a>0$, the bound state has a molecular structure,
with the particles having a large mean separation of order $a$.  

To see that the universal features of an S-wave threshold
resonance are relevant to the $X(3872)$, 
we need only note that its binding energy is small compared to 
the natural energy scale associated with pion
exchange \cite{Braaten:2003he}: $m_\pi^2 / (2M_{*00}) \approx 10$ MeV, 
where $M_{*00}$ is the reduced mass of the two constituents. 
The universal features of the $X(3872)$ were first exploited by
Voloshin to describe its decays into $D^0 \bar D^0 \pi^0$ and
$D^0 \bar D^0 \gamma$, which can proceed through decay of the
constituent $D^{*0}$  or $\bar{D}^{*0}$ \cite{Voloshin:2003nt}.
Universality has also been applied to the
production process $B \to KX$ \cite{Braaten:2004fk,Braaten:2004ai},
to the line shapes of the $X$ \cite{Braaten:2005jj}, and
to decays of $X$ into $J/\psi$ and pions \cite{Braaten:2005ai}.
These applications rely on factorization formulas that separate
the length scale $a$ from all the shorter distance scales of QCD
\cite{Braaten:2005jj}.  The factorization formulas can be derived
using the operator product expansion for a low-energy
effective field theory \cite{Braaten:2006sy}.

Other interpretations of the $X(3872)$ 
besides a charm meson molecule or a charm meson virtual state
have been proposed, including a P-wave charmonium state 
or a tetraquark state. (For a review, see Ref.~\cite{Swanson:2006st}.)
If the charmonium or tetraquark 
models were extended to include the coupling of
the $X$ to $D^{*0} \bar D^0$ and $D^0 \bar{D}^{*0}$
scattering  states, the universal features of an S-wave threshold
resonance imply that the tuning 
of the binding energy to the threshold region would transform 
the state into a charm meson molecule or a virtual state of charm mesons. 
Any model of the $X(3872)$ that does not take into account its 
strong coupling to charm meson scattering states 
should not be taken seriously.  

Given that the quantum numbers of the $X(3872)$ are $1^{++}$, 
the measured mass $M_X$ in Eq.~(\ref{MX-CLEO})
implies unambiguously that $X$ must be either a 
charm meson molecule or a virtual state of charm mesons.
The remaining challenge is to discriminate between these two possibilities.
If the $X$ was sufficiently narrow, there would be
clear qualitative differences in its line shapes
between these two possibilities.
We first consider the $D^0 \bar D^0 \pi^0$ decay mode,
which has a contribution from the decay of a constituent $D^{*0}$.
If the $X$ was a charm meson molecule, its line shape
in $D^0 \bar D^0 \pi^0$
would consist of a Breit-Wigner resonance below the $D^{*0} \bar D^0$
threshold and a threshold enhancement above the $D^{*0} \bar D^0$
threshold.  If the $X$ was a virtual state, 
there would only be the threshold enhancement above the $D^{*0} \bar D^0$
threshold.  We next consider decay modes that have no contributions 
from the decay of a constituent $D^{*0}$, such as $J/\psi \, \pi^+ \pi^-$.
If the $X$ was a charm meson molecule, its line shape in such a 
decay mode would be a Breit-Wigner resonance below the 
$D^{*0} \bar D^0$ threshold.  If the $X$ was a virtual state, 
there would only be a cusp at the $D^{*0} \bar D^0$ threshold.
The possibility of interpreting the $ X(3872)$ as a cusp at the 
$D^{*0} \bar D^0$ threshold has been suggested by Bugg \cite{Bugg:2004rk}.
Increasing the width of the $X$ provides
additional smearing of the line shapes.  This makes the qualitative 
difference between the line shapes of a charm meson molecule
and a virtual state less dramatic.
To discriminate between these two possibilities therefore requires 
a quantitative analysis.

There have been two recent analyses of data on $B \to K+X$ 
that shed light on the issue of whether the $X$ is a bound state 
or a virtual state.  Hanhart et al.~\cite{Hanhart:2007yq} 
analyzed the  data on
$B^+ \to K^+ + J/\psi \, \pi^+ \pi^-$ and 
$B^+ \to K^+ + D^0 \bar D^0 \pi^0$ from the Belle and Babar 
Collaborations using a model for the scattering amplitude 
in the $D^{*0} \bar D^0 + D^0 \bar D^{*0}$ channel
that is a generalization of the Flatt\'e parametrization 
for a near-threshold resonance.
They concluded that the $D^0 \bar D^0 \pi^0$ 
threshold enhancement observed by the Belle and Babar 
Collaborations is compatible with the $X(3872)$ 
only if the $X$ is a virtual state.
One flaw in the analysis of Ref.~\cite{Hanhart:2007yq} is that
it did not take into account the width of the constituent $D^{*0}$.  
They also assumed incorrectly that a bound state 
below the  $D^{*0} \bar D^0$ threshold would not decay 
into $D^0 \bar D^0 \pi^0$.

In Ref.~\cite{Braaten:2007dw}, we derived the line shapes of the 
$X(3872)$ near the $D^{*0} \bar D^0$ threshold from the 
assumption of an S-wave resonance in the
neutral charm meson channel $D^{*0} \bar D^0 + D^0 \bar D^{*0}$. 
We developed expressions for the line shapes that take into account 
the width of the $D^{*0}$ meson and the inelastic scattering channels 
of the charm mesons. An analysis of the data on 
$B^+ \to K^+ + J/\psi \, \pi^+ \pi^-$ and
$B^+ \to K^+ + D^0 \bar D^0 \pi^0$ from the Belle Collaboration 
indicated that the data preferred the $X(3872)$ to be a bound state 
but a virtual state was not excluded. The most important lesson 
of the analyses of Refs.~\cite{Hanhart:2007yq} and 
\cite{Braaten:2007dw} is that the measured difference between 
the masses of the $X(3872)$ in the $J/\psi \, \pi^+ \pi^-$ 
and $D^0 \bar D^0 \pi^0$ decay channels is consistent with it 
being a charm meson molecule or a virtual state of charm mesons.

In this paper, we generalize the results of 
Ref.~\cite{Braaten:2007dw} for the line shapes of the $X(3872)$
to take into account the resonant coupling between the neutral charm 
meson channel and the charged charm meson channel
$D^{*+} D^- + D^+ D^{*-}$. 
In Sec.~\ref{sec:widths}, we summarize the results of 
Ref.~\cite{Braaten:2007dw} for the energy-dependent widths 
of virtual $D^*$ mesons.  In Ref.~\cite{Braaten:2007dw},
we developed an expression for the resonant scattering amplitude 
for the neutral charm meson channel that takes into account 
the $D^{*0}$ width and inelastic charm meson scattering channels.
In Sec.~\ref{sec:DDscat}, we extend that result to the three 
scattering amplitudes for the resonantly coupled 
neutral and charged charm meson channels.
In Ref.~\cite{Braaten:2007dw}, we derived factorization formulas 
for the line shapes of $X(3872)$ in the decays $B \to K + X$
that take into account the resonance in the neutral charm meson 
channel.  In Section~\ref{sec:lineshape}, we extend those results 
to take into account the resonant coupling to the 
charged charm meson channel. 
In Sec.~\ref{sec:summary}, we summarize our results.

\section{Masses and $\bm{D^*}$ Widths}
\label{sec:widths}

When we consider the decays of the $D^*$ mesons,
there are particles with six different masses that must be considered.
We therefore introduce concise notation for the masses
of the charm mesons and the pions.
We denote the masses of the spin-0 charm mesons
$D^0$ and $D^+$ by $M_0$ and $M_1$, respectively.
We denote the masses of the spin-1 charm mesons
$D^{*0}$ and $D^{*+}$ by $M_{*0}$ and $M_{*1}$, respectively.
We denote the masses of the pions
$\pi^0$ and $\pi^+$ by $m_0$ and $m_1$, respectively.
(The numerical subscript is the absolute value of the
electric charge of the meson.)
The pion mass scale corresponding to either $m_0$ or $m_1$ 
will be denoted by $m_\pi$.
The result of a recent precision measurement of the $D^0$ mass by the
CLEO Collaboration is $M_0 = 1864.85 \pm 0.18$ MeV,
where we have combined the errors in quadrature \cite{Cawlfield:2007dw}.
We use the PDG values for the other masses \cite{Yao:2006px}.
The errors on the pion masses are negligible compared to those
on the charm meson masses.
Some of the differences between the charm meson masses
have errors that are significantly smaller than the
errors in the masses themselves.

We also introduce concise notations for simple
combinations of the masses.
We denote the reduced mass of a spin-1 charm meson
and a spin-0 charm meson by
\begin{eqnarray}
M_{*ij} &=& \frac{M_{*i}M_j}{M_{*i}+M_j} .
\end{eqnarray}
We denote the reduced mass of a pion
and a spin-0 charm meson by
\begin{eqnarray}
m_{ij} &=& \frac{m_i M_j}{m_i+M_j} .
\end{eqnarray}
We denote the differences
between the $D^*$ masses and $D \pi$ thresholds by
\begin{equation}
\delta_{ijk} = M_{*i} - M_j - m_k  .
\end{equation}
The differences between the $D^*$ masses and the thresholds
for $D \pi$ states with the same electric charge are
\begin{subequations}
\begin{eqnarray}
\delta_{000}
&=&  7.14 \pm 0.07~{\rm MeV} ,
\label{delta000}
\\
\delta_{011}
&=&  -2.23 \pm 0.12~{\rm MeV} ,
\label{delta011}
\\
\delta_{101}
&=&  5.85 \pm 0.01~{\rm MeV}  ,
\label{delta101}
\\
\delta_{110}
&=&  5.66 \pm 0.10~{\rm MeV}  .
\label{delta110}
\end{eqnarray}
\end{subequations}
The isospin splittings between the charm meson masses are
$M_1 - M_0 \approx 4.8$ MeV and $M_{*1} - M_{*0} \approx 3.3$ MeV.
The energy splitting $\nu = (M_{*1} + M_1) - (M_{*0} + M_0)$
between the $D^{*+} D^{-}$ and $D^{*0} \bar D^0$ thresholds is
\begin{eqnarray}
\nu = 8.08 \pm 0.12~{\rm MeV}  .
\label{nu}
\end{eqnarray}

A phenomenological analysis of the decays of the $D^{*0}$ and $D^{*+}$ 
was presented in Ref.~\cite{Braaten:2007dw}.  We summarize here the results 
of that analysis, which was based on chiral symmetry and isospin
symmetry.  The PDG value for the total width of the $D^{*+}$ is
$\Gamma[D^{*+}] = 96 \pm 22$ keV \cite{Yao:2006px}.  Using the PDG values 
for the branching fractions for $D^{*+}$ decays, we obtain 
measured values for the partial widths for $D^{*+}$ decays:
\begin{subequations}
\begin{eqnarray}
\Gamma[D^{*+} \to D^0 \pi^+] &=&  65.0 \pm 14.9 \  {\rm keV} ,
\label{GamD*++0:N}
\\
\Gamma[D^{*+} \to D^+ \pi^0] &=&  29.5 \pm 6.8 \  {\rm keV} ,
\label{GamD*+0+:N}
\\
\Gamma[D^{*+} \to D^+ \gamma\ ] &=& 1.5 \pm 0.5 \  {\rm keV} .
\label{GamD*++gam:N}
\end{eqnarray}
\end{subequations}
Using isospin symmetry and the PDG values for the
branching fractions for $D^{*0}$ decays, 
we obtain predictions for the partial widths for $D^{*0}$ decays:
\begin{subequations}
\begin{eqnarray}
\Gamma[D^{*0} \to D^0 \pi^0] &=& 40.5 \pm 9.3 \  {\rm keV} ,
\label{GamD*000:N}
\\
\Gamma[D^{*0} \to D^0 \gamma ] &=&  25.0 \pm 6.2 \  {\rm keV} .
\label{GamD*00gam}
\end{eqnarray}
\end{subequations}
The prediction for the total width of the $D^{*0}$ is
$\Gamma[D^{*0}] = 65.5 \pm 15.4$ keV.

\begin{figure}[t]
\includegraphics[width=12cm]{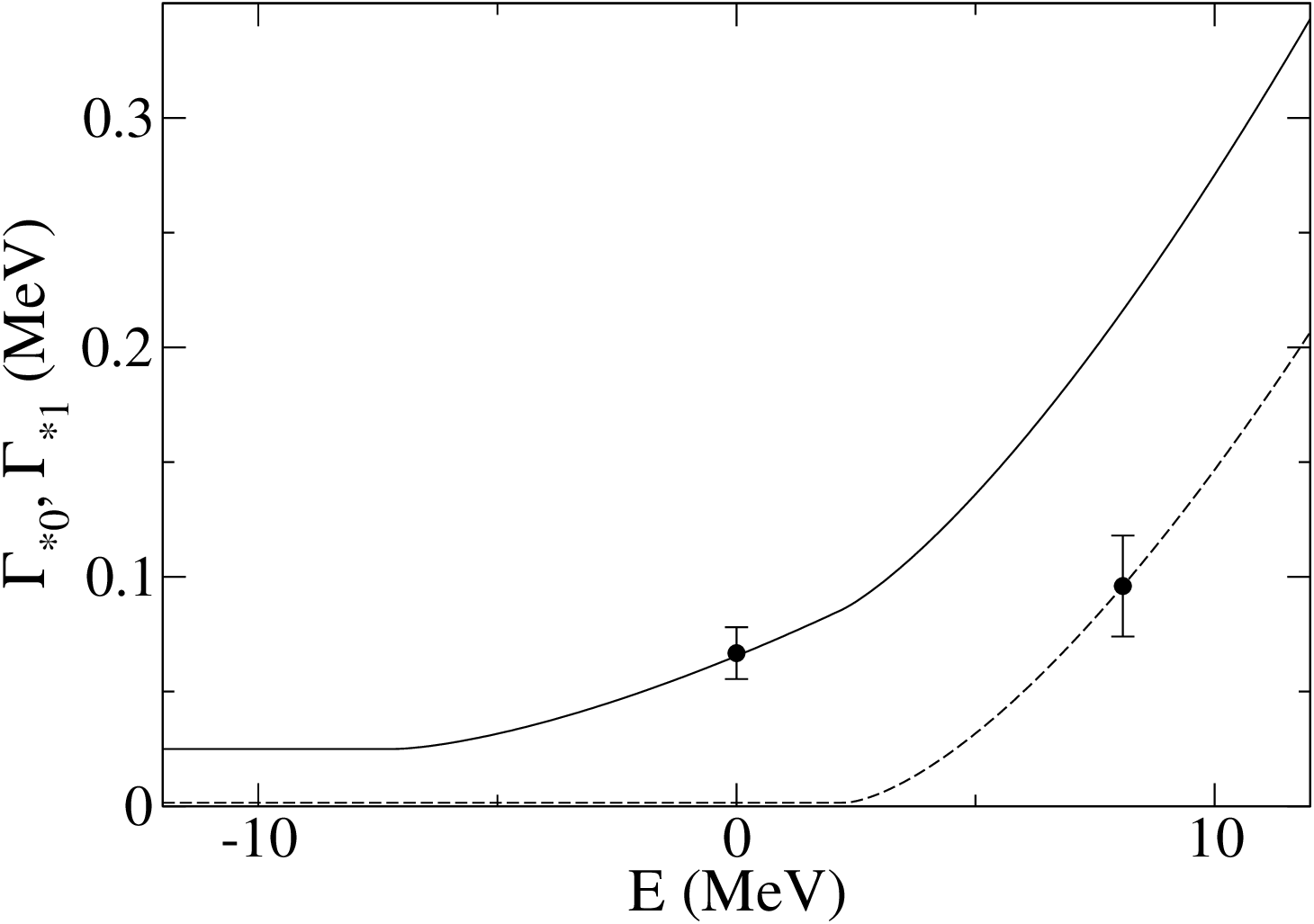}
\caption{
The energy-dependent widths $\Gamma_{*0}(E)$ and $\Gamma_{*1}(E - \nu)$  
for a virtual $D^{*0}$ with energy $M_{*0} + E$ and a virtual
$D^{*+}$ with energy $M_{*1} + E - \nu$, respectively, as functions of $E$.
The points with error bars at $E=0$ and $E=\nu$
indicate the central values and uncertainties of the physical widths
of $D^{*0}$ and $D^{*+}$, respectively.
\label{fig:Gam*}}
\end{figure}

The decay rates for $D^* \to D \pi$ are fairly sensitive 
to the mass of the $D^*$, since they scale like the 3/2 power of the
energy difference between the $D^*$ mass and the $D \pi$ threshold.
A virtual $D^{*0}$ (or $D^{*+}$)
with energy $M_{*0}+E$ (or $M_{*1}+E$) can be considered as a $D^*$ whose
rest energy differs from its physical mass by the energy $E$.
The width of the virtual particle varies with $E$.
We denote the energy-dependent widths of the $D^{*+}$ and $D^{*0}$
by $\Gamma_{*1}(E)$ and $\Gamma_{*0}(E)$, respectively.
If $|E|$ is small compared to $m_\pi$, 
these energy-dependent widths can be obtained simply by scaling 
the physical partial widths for the decays $D^*\to D \pi$:
\begin{subequations}
\begin{eqnarray}
\Gamma_{*0}(E) &=&
\Gamma[D^{*0} \to D^0 \gamma]
+ \Gamma[D^{*0} \to D^0 \pi^0]
\big[  \left[ (\delta_{000} + E)/\delta_{000} \right]^{3/2}
\theta(\delta_{000} + E)
\nonumber \\
&& \hspace{3cm}
+ 2 \left( m_{11}/m_{00} \right)^{5/2}
\left[ (\delta_{011} + E)/\delta_{000} \right]^{3/2}
\theta(\delta_{011} + E) \big] ,
\label{GamD*0-E}
\\
\Gamma_{*1}(E) &=&
\Gamma[D^{*+} \to D^+ \gamma] +
\Gamma[D^{*+} \to D^+ \pi^0]
\left[ (\delta_{110} + E)/\delta_{110}\right]^{3/2}
\theta(\delta_{110} + E)
\nonumber \\
&& \hspace{3cm}
+ \Gamma[D^{*+} \to D^0 \pi^+]
\left[ (\delta_{101} + E)/\delta_{101} \right]^{3/2}
\theta(\delta_{101} + E) .
\label{GamD*+-E}
\end{eqnarray}
\label{GamD*-E}
\end{subequations}
We ignore any energy dependence of the decay widths into $D \gamma$,
because the photon energy and the phase space
for the decays $D^*\to D \gamma$
do not vary significantly in the $D^* \bar D$ threshold region.
In Fig.~\ref{fig:Gam*}, we plot the energy-dependent widths 
$\Gamma_{*0}(E)$ and $\Gamma_{*1}(E - \nu)$ as functions of $E$. 
The offset $\nu \approx 8.1$ MeV in $\Gamma_{*1}(E - \nu)$
was chosen so that $\Gamma_{*0}(E)$ and $\Gamma_{*1}(E - \nu)$
are the relevant widths for a $D^* \bar D$ system consisting of
\ $\bar D$ and a $D^*$ with total energy $E$ relative to the  
$D^{*0} \bar D^0$ threshold.  Thus $\Gamma_{*0}(E)$ reduces to
$\Gamma[D^{*0}]$ at $E =0$ and $\Gamma_{*1}(E - \nu)$
reduces to $\Gamma[D^{*+}]$ at $E= \nu$.
The physical widths $\Gamma[D^{*0}]$ and  $\Gamma[D^{*+}]$
are shown in Fig.~\ref{fig:Gam*} as data points with error bars.
At the $D^{*0} \bar D^0$ threshold, the energy-dependent width 
of the $D^{*+}$ is $\Gamma_{*1}(- \nu) \approx 1.5$ MeV.

The individual terms in Eqs.~(\ref{GamD*-E})
have obvious interpretations as energy-dependent partial widths
for decays of $D^{*+}$ and $D^{*0}$.  We can define energy-dependent 
branching fractions by dividing these terms by $\Gamma_{*1}(E)$ 
or $\Gamma_{*0}(E)$.
For example, the energy-dependent branching fractions
for $D^{*0} \to D^0 \pi^0$ and $D^{*+} \to D^+ \pi^0$ are
\begin{subequations}
\begin{eqnarray}
{\rm Br}_{000}(E) &=&
\frac{\Gamma[D^{*0} \to D^0 \pi^0]}{\Gamma_{*0}(E)} \,
\left[ (\delta_{000} + E)/\delta_{000} \right]^{3/2}
\theta(\delta_{000} + E) ,
\label{Br000}
\\
{\rm Br}_{110}(E) &=&
\frac{\Gamma[D^{*+} \to D^+ \pi^0]}{\Gamma_{*1}(E)} \,
\left[ (\delta_{110} + E)/\delta_{110} \right]^{3/2}
\theta(\delta_{110} + E) .
\label{Br110}
\end{eqnarray}
\end{subequations}

The standard isospin multiplets for the charm mesons are
$( -D^+, D^0)$, $( \bar D^{0}, D^{-})$, $( -D^{*+}, D^{*0})$,
and $( \bar D^{*0}, D^{*-})$,
where the first and second states are the upper and lower
components of the multiplet, respectively.
The $D^* \bar D$ channels with charge conjugation quantum number
$C=+$ are
\begin{subequations}
\begin{eqnarray}
(D^* \bar D)_+^0 &=&  +\mbox{$\frac{1}{\sqrt{2}}$}
\left( D^{*0} \bar D^0  + D^0 \bar D^{*0} \right) ,
\label{D*Dbar0}
\\
(D^* \bar D)_+^1 &=&  -\mbox{$\frac{1}{\sqrt{2}}$}
\left( D^{*+} D^-  + D^+ D^{*-} \right).
\label{D*Dbar1}
\end{eqnarray}
\label{D*Dbar}
\end{subequations}
The superscript $i$ on $(D^* \bar D)_+^i$ is the 
absolute value of the electric charge of either meson.
We will refer to $(D^* \bar D)_+^0$ and $(D^* \bar D)_+^1$
as the neutral and charged charm meson channels, respectively. 
The channels with isospin quantum numbers $I=0$ and $I=1$
are the antisymmetric and symmetric linear combinations of
these neutral and charged channels, respectively:
\begin{subequations}
\begin{eqnarray}
(D^* \bar D)_+^{I=0} &=&  \mbox{$\frac{1}{\sqrt{2}}$}
\left[ (D^* \bar D)_+^0 - (D^* \bar D)_+^1 \right],
\label{D*Dbar:I=0}
\\
(D^* \bar D)_+^{I=1} &=&  \mbox{$\frac{1}{\sqrt{2}}$}
\left[ (D^* \bar D)_+^0 + (D^* \bar D)_+^1 \right].
\label{D*Dbar:I=1}
\end{eqnarray}
\label{D*Dbar:I}
\end{subequations}

\section{Low-energy $\bm{D^{*} {\bar D}}$ Scattering}
\label{sec:DDscat}

In this section, we discuss the low-energy scattering of charm mesons.
We first summarize the results of  Ref.~\cite{Braaten:2007dw},
which took into account only the neutral channel $(D^* \bar D)_+^0$.
These results should be accurate when the energy $E$ 
is within a few MeV of  the $D^{*0} \bar D^0$ threshold. 
We then extend the region of validity to the entire $D^* \bar D$
threshold region by taking into account the resonant coupling to
the charged channel $(D^* \bar D)_+^1$.

\subsection{Neutral channel only}
\label{sec:DDscat1ch}

We begin by writing down a general expression for the transition 
amplitude for S-wave scattering in the $(D^* \bar D)_+^0$ channel 
that is compatible with  unitary.  
The transition amplitude $\mathcal{A}(E)$ 
for the scattering of nonrelativistically normalized
charm mesons in the channel $(D^* \bar D)_+^0$ 
can be written in the form
\begin{equation}
\mathcal{A}(E) = \frac{2 \pi}{M_{*00}} f(E) ,
\end{equation}
where $f(E)$ is the conventional nonrelativistic scattering amplitude
expressed as a function of the total energy of the charm mesons.
An expression for the scattering amplitude that is 
compatible with unitarity is 
\begin{equation}
f(E)  = \frac{1}{- \gamma + \kappa(E)} ,
\label{f-E}
\end{equation}
where $\kappa(E)= (- 2 M_{*00} E - i \varepsilon)^{1/2}$
and $E$ is the total energy relative to the $D^{*0} \bar D^0$ 
threshold in the center-of-mass frame.
If the inverse scattering length $\gamma$ is complex,
the imaginary part of the scattering amplitude in Eq.~(\ref{f-E}) is
\begin{equation}
{\rm Im} \, f(E)  =
| f(E) |^2 \,
{\rm Im} \left[ \gamma - \kappa(E)  \right] .
\label{ImA-optical2}
\end{equation}

The scattering amplitude $f(E)$ in Eq.~(\ref{f-E}) 
satisfies the constraints of unitarity for a single-channel system 
exactly provided $\gamma$ is a real function of $E$.
For positive real values of the energy $E$,
Eq.~(\ref{ImA-optical2})
is simply the optical theorem for this single-channel system:
\begin{equation}
{\rm Im} \, f(E)  =
|  f(E) |^2
\sqrt{2 M_{*00} E}
\hspace{1cm} (E>0).
\label{ImA-optical:1ch}
\end{equation}
The left side is the imaginary part of the T-matrix element
for elastic scattering in the $(D^* \bar D)_+^0$ channel
multiplied by $M_{*00}/(2 \pi)$.
The right side is the cross section for elastic scattering
multiplied by $(2 M_{*00}E)^{1/2}/(4 \pi)$.
We first consider the case $\gamma > 0$.
In this case, the amplitude $f(E)$
has a pole at a negative value of the energy $E$,
indicating the existence of a stable bound state.
If $\gamma$ varies sufficiently slowly with $E$ 
that it can be approximated by a constant,
the pole is near $E_{\rm pole} \approx - \gamma^2/(2 M_{*00})$
and the binding energy is $\gamma^2/(2 M_{*00})$. 
In addition to the contribution to the imaginary part of $f(E)$
in Eq.~(\ref{ImA-optical:1ch}), there is a delta-function 
contribution at $E= E_{\rm pole}$:
\begin{equation}
{\rm Im} \, f(E)  \approx
\frac{\pi \gamma}{M_{*00}} 
\delta ( E + \gamma^2/(2 M_{*00})) 
\hspace{1cm} (E<0,\gamma>0).
\label{ImA-optical:1chE<0}
\end{equation}
We next consider the case $\gamma < 0$.  In this case,
the pole in the amplitude $f(E)$ is not on the real
$E$ axis, but on the second sheet of the complex variable $E$.
The standard terminology for such a pole is a {\it virtual state}.
The imaginary part of the amplitude is nonzero 
only in the positive $E$ region and is given by 
Eq.~(\ref{ImA-optical:1ch}).

Scattering in the $(D^* \bar D)_+^0$ channel cannot be exactly 
unitary, because the $D^{*0}$ has a nonzero width and because 
the charm mesons have inelastic scattering channels.
The inelastic channels include $D^0 \bar D^0 \pi^0$ 
and $D^0 \bar D^0 \gamma$, which are related to $D^{*0}$ or 
$\bar D^{*0}$ decays, as well as all other
decay modes of $X(3872)$, including $J/\psi \, \pi^+ \pi^-$,
$J/\psi \, \pi^+ \pi^- \pi^0$, and $J/\psi \, \gamma$.
In Ref.~\cite{Braaten:2007dw}, the dominant effects of the $D^{*0}$ width 
and the inelastic scattering channels were taken into account through
simple modifications of the variables $\gamma$ and $\kappa(E)$
in the scattering amplitude $f(E)$ in Eq.~(\ref{f-E}).
The effects of the decays of the constituent $D^{*0}$
or $\bar{D}^{*0}$ were taken into account simply by replacing
the mass $M_{*0}$ that is implicit in the energy $E$ measured
from the $D^{*0} \bar D^0$ threshold by
$M_{*0} - i \Gamma_{*0}(E)/2$,
where $\Gamma_{*0}(E)$ is the energy-dependent width
of the $D^{*0}$ given in Eq.~(\ref{GamD*0-E}).
This changes the energy variable 
$\kappa(E)= (- 2 M_{*00} E - i \varepsilon)^{1/2}$ into
\begin{equation}
\kappa(E)  = \sqrt{- 2 M_{*00} [E + i \Gamma_{*0}(E)/2]} .
\label{kappa-E}
\end{equation}
At the threshold $E=0$, the energy-dependent width $\Gamma_{*0}(E)$
reduces to the physical width $\Gamma[D^{*0}]$.
The expression for $\kappa(E)$ in Eq.~(\ref{kappa-E})
requires a choice of branch cut for the square root.
If $E$ is real, an explicit expression for $\kappa(E)$ 
that corresponds to the appropriate choice of branch cut
can be obtained by using the identity
\begin{eqnarray}
\sqrt{- 2 M [E + i \Gamma/2]}
&=&  \sqrt{M}
\left[ \left( \sqrt{E^2 + \Gamma^2/4} - E \right)^{1/2}
   - i  \left( \sqrt{E^2 + \Gamma^2/4} + E \right)^{1/2}
\right] .
\label{sqrtid}
\end{eqnarray}
In Ref.~\cite{Braaten:2007dw}, the effects of inelastic scattering 
channels for the charm mesons other than $D^0 \bar D^0 \pi^0$ and
$D^0 \bar D^0 \gamma$ were taken into account by replacing
the real parameter $\gamma$ by a complex parameter
with a positive imaginary part.  
The expression for the imaginary part of the amplitude  
$f(E)$ in Eq.~(\ref{ImA-optical2}) can now be interpreted 
as the optical theorem for a multi-channel system consisting 
of $(D^* \bar D)_+^0$ and all the inelastic scattering channels.
The right side can be interpreted as the total cross section 
for scattering in the $(D^* \bar D)_+^0$ channel multiplied by 
$(2 M_{*00}E)^{1/2}/(4 \pi)$. 
The terms proportional to ${\rm Im} \kappa(E)$ and ${\rm Im} \gamma$
are proportional to the elastic and inelastic cross sections, 
respectively. This interpretation requires 
${\rm Im} \gamma > 0$. 

The scattering amplitude $f(E)$ in Eq.~(\ref{f-E}) with $\kappa(E)$
given by Eq.~(\ref{kappa-E}) and a complex parameter $\gamma$
has a pole at an energy $E_{\rm pole}$ that is not on the real axis.
If the difference between $\Gamma_{*0}(E_{\rm pole})$
and $\Gamma_{*0}(0)$ is small compared to $\Gamma[D^{*0}]$,
the pole energy can be approximated by
\begin{equation}
E_{\rm pole} \approx - \frac{\gamma^2}{2 M_{*00}} - i \Gamma[D^{*0}]/2 .
\label{Epole}
\end{equation}
The energy $E_{\rm pole}$ is that of a bound state if 
${\rm Re} \gamma>0$ and that
of a virtual state if ${\rm Re} \gamma < 0$.

\subsection{Coupled neutral and charged channels}
\label{sec:DDscat2ch}

We now generalize the results of Section~\ref{sec:DDscat1ch} 
to the system consisting of the two coupled 
channels $(D^* \bar D)_+^0$ and $(D^* \bar D)_+^1$
defined by Eqs.~(\ref{D*Dbar}).  The amplitudes for transitions between
these channels can be expressed as a $2 \times 2$ matrix
$\mathcal{A}_{ij}(E)$, $i,j \in \{ 0,1 \}$.
We first write down a general expression for the transition 
amplitudes for S-wave scattering in the two channels that is 
compatible with unitarity in this two-channel system.
A convenient way to parametrize these amplitudes
is to express the inverse of the matrix of amplitudes
$\mathcal{A}_{ij}(E)$ in the form
\begin{equation}
\mathcal{A}(E)^{-1} =
\frac{1}{2 \pi}
\left( \begin{array}{cc}
\sqrt{M_{*00}} & 0           \\
0            & \sqrt{M_{*11}}
\end{array} \right)
\left( \begin{array}{cc}
-\gamma_{00} + \kappa(E) & - \gamma_{01} \\
-\gamma_{01}              & -\gamma_{11} + \kappa_1(E)
\end{array} \right)
\left( \begin{array}{cc}
\sqrt{M_{*00}} & 0           \\
0            & \sqrt{M_{*11}}
\end{array} \right) ,
\label{A-inverse}
\end{equation}
where $\kappa(E)= (- 2 M_{*00} E - i \varepsilon)^{1/2}$,
$\kappa_1(E)  = (- 2 M_{*11} (E - \nu) - i \varepsilon)^{1/2}$,
and $E$ is the energy relative to the $D^{*0} \bar D^0$ threshold.
The parametrization of the inverse matrix in Eq.~(\ref{A-inverse})
was chosen so that the analytic expressions for the
entries of $\mathcal{A}_{ij}(E)$ would be as simple as possible.
It is convenient to define a matrix $f_{ij}(E)$
of scattering amplitudes by
\begin{equation}
\mathcal{A}_{ij}(E) = 
\frac{2 \pi}{\sqrt{M_{*ii} M_{*jj}}} 
f_{ij}(E) .
\end{equation}
The entries of the matrix $f_{ij}(E)$ are
\begin{subequations}
\begin{eqnarray}
f_{00}(E) &=&
\left( - \gamma_{00} + \kappa(E)
- \frac{\gamma_{01}^2}{ -\gamma_{11} + \kappa_1(E)} \right)^{-1} ,
\label{f00}
\\
f_{01}(E) &=&
\left( - \gamma_{01} + \frac{[ -\gamma_{00} + \kappa(E) ]
[ -\gamma_{11} + \kappa_1(E) ]}{\gamma_{01}} \right)^{-1},
\\
f_{11}(E) &=&
\left( - \gamma_{11} + \kappa_1(E)
- \frac{\gamma_{01}^2}{ -\gamma_{00} + \kappa(E)} \right)^{-1} .
\end{eqnarray}
\label{fij}
\end{subequations}
If the parameters $\gamma_{00}$, $\gamma_{01}$, and  $\gamma_{11}$
are complex, the imaginary parts of the scattering amplitudes
in Eq.~(\ref{fij}) satisfy
\begin{eqnarray}
\mathrm{Im} \, f_{ij}(E)
&=&
f_{i0}(E) \, f_{j0}^*(E) \,
\mathrm{Im} \left[ \gamma_{00} - \kappa(E) \right]
+
f_{i1}(E) \, f_{j1}^*(E) \,
\mathrm{Im} \left[ \gamma_{11} - \kappa_1(E) \right]
\nonumber
\\
&&
+ \big( f_{i0}(E) \, f_{j1}^*(E) 
+ f_{i1}(E) \, f_{j0}^*(E) \big)
\mathrm{Im} \left[ \gamma_{01} \right] .
\label{ImAij}
\end{eqnarray}
Since $\mathrm{Im} f_{ij}(E)$ is real,
an alternative form for this unitarity equation can be 
obtained by taking the complex conjugate of the right side.

The amplitudes $f_{ij}(E)$ in Eqs.~(\ref{fij}) satisfy the constraints 
of unitarity for this two-channel system exactly if 
$\gamma_{00}$, $\gamma_{01}$, and $\gamma_{11}$
are all real functions of $E$.
For positive real values of the energy $E$, the expressions  
in Eqs.~(\ref{ImAij}) for the imaginary parts of the amplitudes 
$f_{00}(E)$ and $f_{11}(E)$
are just the optical theorems for this two-channel system:
\begin{subequations}
\begin{eqnarray}
{\rm Im} \, f_{00}(E)  &=&
| f_{00}(E) |^2 \sqrt{2 M_{*00} E}
+ | f_{01}(E) |^2 \sqrt{2 M_{*11} (E-\nu)} \, \theta(E-\nu),
\label{ImA00:E>0}
\\
{\rm Im} \, f_{11}(E)  &=&
| f_{01}(E) |^2 \sqrt{2 M_{*00} E}
+ | f_{11}(E) |^2 \sqrt{2 M_{*11} (E-\nu)} \, \theta(E-\nu) .
\label{ImA11:E>0}
\end{eqnarray}
\end{subequations}
The left sides of Eqs.~(\ref{ImA00:E>0}) and (\ref{ImA11:E>0}) 
are proportional to the imaginary parts of the T-matrix elements
for elastic scattering in the $(D^* \bar D)_+^0$ 
and $(D^* \bar D)_+^1$ channels, respectively.
The first and second terms on the right side of each equation 
are proportional to the cross sections
for scattering into the $(D^* \bar D)_+^0$ and $(D^* \bar D)_+^1$ channels, 
respectively.  In the region $E<0$, the imaginary parts
of $f_{00}(E)$ and $f_{11}(E)$ may also have delta function
contributions analogous to the one in Eq.~(\ref{ImA-optical:1chE<0}).

There are two limits in which
the amplitudes $f_{01}(E)$ and $f_{11}(E)$
go to 0 and $f_{00}(E)$ reduces to the single-channel
amplitude in Eq.~(\ref{f-E}).
The first limit is $\nu \to + \infty$, which corresponds to increasing
the energy gap between the two thresholds.
In this case, $f_{00}(E)$ reduces to Eq.~(\ref{f-E})
with $\gamma = \gamma_{00}$. The second limit is
$\gamma_{01}, \gamma_{11} \to \infty$ with $\gamma_{01}^2/\gamma_{11}$
fixed, which corresponds to decreasing the interaction strength
between the two channels.  In this case, $f_{00}(E)$
again reduces to Eq.~(\ref{f-E}) but with
$\gamma = \gamma_{00}-\gamma_{01}^2/\gamma_{11}$.

Scattering in the $(D^* \bar D)_+^0$ and $(D^* \bar D)_+^1$ 
channels cannot be exactly unitary, because the $D^{*0}$ and 
$D^{*+}$ have nonzero widths and because 
the charm mesons have inelastic scattering channels.
The inelastic channels include $D \bar D \pi$ and $D \bar D \gamma$, 
which are related to $D^*$ or $\bar D^*$ decays, as well as 
all the decay modes of $X(3872)$, which include $J/\psi \, \pi^+ \pi^-$,
$J/\psi \, \pi^+ \pi^- \pi^0$, and $J/\psi \, \gamma$.
The effects of decays of $D^{*0}$ and $\bar D^{*0}$ can be
taken into account by replacing $\kappa(E)$ in the amplitudes 
in Eqs.~(\ref{fij}) by the expression in Eq.~(\ref{kappa-E}).
Similarly the effects of decays of $D^{*+}$ and $D^{*-}$
can be taken into account by replacing $\kappa_1(E)$ 
in the amplitudes in Eqs.~(\ref{fij}) by
\begin{equation}
\kappa_1(E)  =
\sqrt{- 2 M_{*11} [E - \nu + i \Gamma_{*1}(E - \nu)/2]} .
\label{kappa1-E}
\end{equation}
At the $D^{*+} D^-$ threshold $E=\nu$, 
the energy-dependent width $\Gamma_{*1}(E - \nu)$
reduces to the physical width $\Gamma[D^{*+}]$.
If $E$ is real, an explicit expression for $\kappa_1(E)$ 
that corresponds to the appropriate choice of the square-root branch cut
in Eq.~(\ref{kappa1-E}) can be obtained by using the identity in 
Eq.~(\ref{sqrtid}).  The effects of inelastic scattering channels
other than $D \bar D \pi$ and $D \bar D \gamma$ can be taken into account 
by replacing the real parameters $\gamma_{00}$, $\gamma_{01}$, 
and $\gamma_{11}$ by complex parameters with positive imaginary parts.
The expression in Eq.~(\ref{ImAij}) for the imaginary part of the 
amplitude $f_{ii}(E)$ can now be interpreted 
as the optical theorem for the multi-channel system consisting of
$(D^* \bar D)_+^0$, $(D^* \bar D)_+^1$, and all the inelastic 
scattering channels.
The right side can be interpreted as the total cross section 
for scattering in the $(D^* \bar D)_+^i$ channel
multiplied by $(2 M_{*ii}E)^{1/2}/(4 \pi)$.
The terms proportional to ${\rm Im} \kappa(E)$ and 
${\rm Im} \kappa_1(E)$ are the cross sections for scattering into
the $(D^* \bar D)_+^0$ and $(D^* \bar D)_+^1$ channels. 
The terms proportional to ${\rm Im} \gamma_{00}$, 
${\rm Im} \gamma_{01}$, and ${\rm Im} \gamma_{11}$ 
give the remaining inelastic cross sections.  This interpretation 
requires ${\rm Im} \gamma_{00} > 0$ and ${\rm Im} \gamma_{11} > 0$.

\subsection{Constraints from isospin symmetry}

We now proceed to exploit the approximate isospin symmetry
of QCD.  Deviations from isospin symmetry can be treated
as small perturbations except at low energies that are comparable
to the isospin splittings between hadron masses,
which in the case of charm hadrons are less than 5 MeV.
In strong interaction processes, isospin-symmetry-violating effects
come primarily from hadron mass differences.  Exact isospin symmetry
would require the masses of the charged charm mesons to be equal
to those of their neutral counterparts, which implies $\nu = 0$
and $M_{*11} = M_{*00}$.  It would also require the inverse matrix 
of amplitudes in Eq.~(\ref{A-inverse}) to be diagonal 
in the isospin basis.  These conditions can be expressed as
\begin{equation}
U
\left( \begin{array}{cc}
-\gamma_{00} + \kappa(E) & - \gamma_{01} \\
-\gamma_{01}              & -\gamma_{11} + \kappa(E)
\end{array} \right)
U^\dagger
= \left( \begin{array}{cc}
-\gamma_{0} + \kappa(E) & 0                   \\
0                   & -\gamma_{1} + \kappa(E)
\end{array} \right)
,
\label{UAUdag}
\end{equation}
where $\gamma_{0}$ and $\gamma_{1}$ are the inverse scattering lengths
in the isospin-symmetry limit
for the $I=0$ and $I=1$ channels, respectively, and $U$ is the
unitary matrix associated with the transformation between the
charged/neutral basis in Eqs.~(\ref{D*Dbar}) and the isospin basis
in Eqs.~(\ref{D*Dbar:I}).  The conditions in Eq.~(\ref{UAUdag})
imply
\begin{subequations}
\begin{eqnarray}
\gamma_{00}  &=& (\gamma_1 + \gamma_0)/2,
\\
\gamma_{01}  &=& (\gamma_1 - \gamma_0)/2,
\\
\gamma_{11}  &=& (\gamma_1 + \gamma_0)/2.
\end{eqnarray}
\label{gammaij-gammak}
\end{subequations}
The constraints on the amplitudes $f_{ij}(E)$ from the approximate
isospin symmetry of QCD are obtained by inserting these values
for the parameters into Eqs.~(\ref{fij}).

In terms of the parameters $\gamma_0$ and $\gamma_1$, 
the scattering amplitudes in Eqs.~(\ref{fij}) reduce to
\begin{subequations}
\begin{eqnarray}
f_{00}(E) &=&
\frac{- (\gamma_0 + \gamma_1) + 2 \kappa_1(E)}{D(E)} ,
\label{f00-E}
\\
f_{01}(E) &=&
\frac{\gamma_1 - \gamma_0}{D(E)} ,
\label{f01-E}
\\
f_{11}(E) &=&
\frac{- (\gamma_0 + \gamma_1) + 2 \kappa(E)}{D(E)} ,
\label{f11-E}
\end{eqnarray}
\label{fij-E}
\end{subequations}
where the denominator is 
\begin{equation}
D(E) = 2 \gamma_1 \gamma_0 
- (\gamma_1 + \gamma_0) [\kappa(E) +  \kappa_1(E)]
+ 2 \kappa_1(E) \kappa(E) .
\label{D-E}
\end{equation}
The unitarity conditions in Eq.~(\ref{ImAij}) can be written
\begin{eqnarray}
\mathrm{Im} \, f_{ij}(E)
&=&
f_{i0}(E) \, f_{j0}^*(E) \,
\mathrm{Im} \left[ \gamma_1 + \gamma_0 - 2 \kappa(E) \right]/2
\nonumber
\\
&& +
f_{i1}(E) \, f_{j1}^*(E) \,
\mathrm{Im} \left[ \gamma_1 + \gamma_0 - 2 \kappa_1(E) \right]/2
\nonumber
\\
&& +
\big( f_{i0}(E) \, f_{j1}^*(E)
+ f_{i1}(E) \, f_{j0}^*(E) \big)
\mathrm{Im} \left[ \gamma_1 - \gamma_0 \right]/2 .
\label{Imfij-I}
\end{eqnarray}

If there is a bound state or virtual state near the 
$D^{*0} \bar D^0$ threshold with complex energy 
$E_{\rm pole}$, the denominator $D(E)$ given in Eq.~(\ref{D-E})
vanishes at that energy.
If we define a variable $\gamma$ by
\begin{equation}
\gamma = \kappa(E_{\rm pole}) ,
\label{gamma-def}
\end{equation}
the equation $D(E_{\rm pole}) = 0$ can be expressed as
\begin{equation}
\gamma \kappa_1(E_{\rm pole}) - \tfrac{1}{2}(\gamma_1 + \gamma_0) 
\left[ \gamma +  \kappa_1(E_{\rm pole}) \right] + \gamma_1 \gamma_0 = 0.
\label{gamma-exact}
\end{equation}
The variable $\gamma$ can be identified with the inverse scattering 
length introduced in Eq.~(\ref{f-E}).
The energy $E_{\rm pole}$ is given approximately by Eq.~(\ref{Epole}).
If we neglect the small difference between $\kappa_1(E_{\rm pole})$ and
$\kappa_1(0)$, one can obtain an approximate solution 
of Eq.~(\ref{gamma-exact}) for $\gamma_0$
in terms of $\gamma_1$ and $\gamma$:
\begin{equation}
\gamma_0 \approx
\frac{\gamma_1 \kappa_1(0) + \gamma_1 \gamma - 2 \kappa_1(0) \gamma}
     {2 \gamma_1 - \kappa_1(0) - \gamma}.
\label{gamma0-approx}
\end{equation}

If the energy $E$ is within a few MeV of the $D^{*0} \bar D^0$ threshold,
the scattering amplitudes in Eqs.~(\ref{fij-E}) can be simplified.
If the small diffrence between $\kappa_1(E)$ and $\kappa_1(0)$ 
is  neglected, the denominator $D(E)$ given in 
Eq.~(\ref{D-E}) reduces to
\begin{equation}
D(E) \approx 
- \left[ \gamma_1 + \gamma_0 - 2 \kappa_1(0) \right]
\left[ - \gamma + \kappa(E) \right] .
\label{D:smallE}
\end{equation}
In the numerators, $\kappa(E)$ and $\gamma$ can be neglected 
compared to $\kappa_1(0)$, $\gamma_0$, and $\gamma_1$.
The scattering amplitudes then reduce to
\begin{subequations}
\begin{eqnarray}
f_{00}(E) &\approx& f(E) ,
\\
f_{10}(E) &\approx& 
\frac{\gamma_0 - \gamma_1}
    {\gamma_1 + \gamma_0 - 2 \kappa_1(0)} f(E) ,
\\
f_{11}(E) &\approx& 
\frac{\gamma_1 + \gamma_0}
    {\gamma_1 + \gamma_0 - 2 \kappa_1(0)} f(E) ,
\end{eqnarray}
\label{fij-smallE}
\end{subequations}
where $f(E)$ is the single-channel scattering amplitude 
in Eq.~(\ref{f-E}).
If $\gamma$ is neglected compared to $\kappa_1(0)$ and $\gamma_1$,
the expression for $\gamma_0$ in Eq.~(\ref{gamma0-approx})
reduces to 
\begin{equation}
\gamma_0 \approx
\frac{\gamma_1 \kappa_1(0)}{2 \gamma_1 - \kappa_1(0)} .
\end{equation}
Using this expression to eliminate $\gamma_0$ in favor of 
$\gamma_1$, the coefficient of $f(E)$ in the amplitudes 
$f_{ij}(E)$ in Eqs.~(\ref{fij-smallE}) can be 
factored into a term that depends on the channel $i$ and a term 
that depends on the channel $j$:
\begin{equation}
f_{ij}(E) \approx c_i \, f(E) \, c_j,
\label{fij:factor}
\end{equation}
where the coefficients $c_i$ are given by
\begin{subequations}
\begin{eqnarray}
c_0 &=& 1,
\\
c_1 &=&  - \frac{\gamma_1}{\gamma_1 - \kappa_1(0)}.
\label{c1-factor}
\end{eqnarray}
\label{cI-factor}
\end{subequations}

The values of the two independent parameters
$\gamma_0$ and $\gamma_1$ could be calculated using 
potential models for heavy mesons with pion-exchange interactions.  
As pointed out by Tornqvist in 1993,
these models indicate that there should be
$D^* \bar D$ bound states near threshold in several $I=0$ channels,
including the S-wave $1^{++}$ channel, but
not in any of the $I=1$ channels \cite{Tornqvist:1993ng}.
Tornqvist could not predict whether the $1^{++}$ state was just 
barely bound or not quite bound, because
his results depended on an ultraviolet cutoff whose value was
estimated to be the same as the corresponding ultraviolet cutoff 
for the two-nucleon system \cite{Tornqvist:1993ng}.
He also could not predict whether the state
would be closer to the $D^{*0} \bar D^0$ threshold or the
$D^{*+} D^-$ threshold, because his calculations were carried out 
in the isospin symmetry limit.
With the discovery of the $X(3872)$, the ambiguity associated 
with the ultraviolet cutoff can be removed by using the observed 
binding energy of the $X(3872)$ to tune the value of the 
ultraviolet cutoff.
One can then use the meson potential model to predict the 
binding energies of other heavy meson molecules
in both the charm sector and the bottom sector \cite{Swanson:2006st}.

In the absence of explicit calculations of the parameters 
$\gamma_0$ and $\gamma_1$, one can still use results of the
meson potential model calculations in Ref.~\cite{Tornqvist:1993ng}
to get some idea of the likely values of these parameters. 
The bound state near threshold with $I=0$ and $J^{PC}=1^{++}$ 
arises from the effects of coupled S-wave and D-wave channels.
In the S-wave channel, the pion-exchange potential is not 
deep enough to give a bound state.  
The D-wave interaction provides just enough additional attraction 
to obtain a bound state very near threshold.
Thus we expect $|\gamma_0|$ to be significantly smaller 
than the natural scale $m_\pi$.
The sign of $\gamma_0$ could be either positive or negative.
The meson potential model calculations in 
Ref.~\cite{Tornqvist:1993ng} indicate that there is no bound state 
with $I=1$ and $J^{PC}=1^{++}$. 
The pion-exchange potential has the opposite sign as in the 
$I=0$ case, so it is repulsive.
We therefore expect $\gamma_1$ to be positive and comparable to 
or larger than the natural scale $m_\pi$.  In particular,
$\gamma_1$ should be much larger than $|\gamma_0|$.
Given an estimate of $\gamma_1$, an estimate of $\gamma_0$ is actually
superfluous because it can be determined using Eq.~(\ref{gamma0-approx}).

The scattering amplitudes $f_{ij}(E)$ in Eqs.~(\ref{fij-E}) 
simplify if the parameter $\gamma_1$ is assumed to be large 
compared to $\kappa_1(0)$.  The denominator $D(E)$ given in 
Eq.~(\ref{D-E}) reduces to
\begin{equation}
D(E) \approx 
- \gamma_1 \left[ -2 \gamma_0 + \kappa(E) + \kappa_1(E) \right] .
\label{D:largegam1}
\end{equation}
The scattering amplitudes reduce to
\begin{equation}
f_{ij}(E) \approx
\frac{1}{-2 \gamma_0 + \kappa_1(E) + \kappa(E)} 
\left( \begin{array}{cc}
           \ 1 &  -1 \\
            -1 & \ 1
\end{array} \right)_{\! \! ij} .
\label{fij:largegam1}
\end{equation}
The matrix projects onto the $I=0$ channel.
The denominator in Eq.~(\ref{D:largegam1}) vanishes at $E_{\rm pole}$.
If the small difference between $\kappa_1(E_{\rm pole})$ 
and $\kappa_1(0)$ is neglected, we get an approximate expression 
for $\gamma_0$ in terms of the variable $\gamma$ defined by 
Eq.~(\ref{gamma-def}):
\begin{equation}
\gamma_0 \approx \frac{\kappa_1(0) + \gamma}{2}.
\label{gamma0-approx2}
\end{equation}

In Ref.~\cite{Hanhart:2007yq}, the authors analyzed data 
from the Belle and Babar Collaborations on the 
energy distributions of $J/\psi \, \pi^+ \pi^-$
and $D^0 \bar D^0 \pi^0$ near the $X(3872)$
resonance produced by the decay $B^+ \to K^+ + X$.
Their model for the $(D^* \bar D)_+^0$ elastic 
scattering amplitude $f(E)$ is a generalization of the 
Flatt\'e parametrization for a near-threshold resonance \cite{Flatte}:
\begin{equation}
f_{\rm HKKN}(E) =
\frac{1}{-(2/g)[E - E_f + i \Gamma(E)/2] + \kappa_1(E) + \kappa(E)} ,
\label{f:HKKN}
\end{equation}
where $\kappa(E) = (-2 M_{*00}E - i \varepsilon)^{1/2}$
and $\kappa_1(E) = (-2 M_{*00}(E - \nu) - i \varepsilon)^{1/2}$.
The function $\Gamma(E)$ is determined 
up to normalization factors by the decays of $X(3872)$.
The other adjustable parameters are $g$ and $E_f$.
In Ref.~\cite{Hanhart:2007yq}, the authors found that their 
fits had a scaling behavior that made it impossible to determine 
unique values of the parameters.  As pointed out in
Ref.~\cite{Braaten:2006sy}, the scaling behavior simply indicates
that their fits were insensitive to the term $-(2/g)E$
in the denominator in Eq.~(\ref{f:HKKN}).  If this term is deleted
and if $-(2/g)E_f$ and $\Gamma(E)/g$ are identified with the real 
and imaginary parts of $2 \gamma_0$, the scattering amplitude 
in Eq.~(\ref{f:HKKN}) reduces to the amplitude $f_{00}(E)$
in Eq.~(\ref{fij:largegam1}), except that it does not take 
into account the effects of the $D^*$ widths.  
As pointed out in Ref.~\cite{Braaten:2007dw}, the $D^{*0}$ width can 
be taken into account by replacing $\kappa(E)$ by the expression in 
Eq.~(\ref{kappa-E}).  Similarly, the $D^{*+}$ width can 
be taken into account by replacing $\kappa_1(E)$ by the expression in 
Eq.~(\ref{kappa1-E}).

\section{Line shapes of $\bm{X(3872)}$}
\label{sec:lineshape}

If a set of particles $C$ has total quantum numbers that are compatible 
with those of the $X(3872)$ resonance and if the total energy $E$
of these particles can be near the $D^{*0} \bar D^0$ threshold, 
then there can be a resonant enhancement in the channel $C$.
The line shape of $X(3872)$ in the channel $C$ is the
differential rate for producing the particles $C$
as a function of their total energy $E$.
In Ref.~\cite{Braaten:2005jj}, it was pointed out that the line shapes
of the $X(3872)$ can be factored into short-distance factors
that are insensitive to $E$ and the inverse scattering length $\gamma$
and a long-distance factor that is determined by $E$ and $\gamma$.
In Ref.~\cite{Braaten:2006sy}, it was shown that the factorization
formulas could be derived using the operator product expansion
for an effective field theory that describes the $c \bar c$
sector of QCD near the $D^{*0} \bar D^0$ threshold.
There is a factorization associated with the creation of the 
charm mesons if all the particles in the initial
state and if the particles in the final state other than 
the resonating particles in $C$
have momenta in the resonance rest frame that are of order
$m_\pi$ or larger.  If $C$ is a short-distance decay mode of $X$, 
there is also a factorization associated with the inelastic scattering
of the charm mesons into the particles in $C$.
A {\it short-distance} decay mode of $X(3872)$ is one 
for which all the particles have momenta that are of order
$m_\pi$ or larger in the resonance rest frame.  
Examples of short-distance decay modes are $J/\psi \, \pi^+ \pi^-$
and $J/\psi \, \pi^+ \pi^- \pi^0$.  An example of a decay mode 
that is not short-distance is $D^0 \bar D^0 \pi^0$.

In this section, we consider the line shapes in the decays
$B \to K + C$, where $C$ is a channel 
that is enhanced by the $X(3872)$ resonance.
We first summarize the results of Ref.~\cite{Braaten:2007dw} 
in which only the neutral channel $(D^* \bar D)_+^0$ 
defined in Eq.~(\ref{D*Dbar0}) was taken into 
account.  These results should be accurate when the energy $E$ 
is within a few MeV of the $D^{*0} \bar D^0$ threshold. 
We then extend the region of validity to the entire
$D^* \bar D$ threshold region by taking into account
the resonant coupling to the charged channel $(D^* \bar D)_+^1$
defined in Eq.~(\ref{D*Dbar1}).

\subsection{Neutral channel only}
\label{sec:lineshape1}

Expressions for the line shapes of the $X(3872)$ that take into 
account the $D^{*0}$ width and inelastic scattering in the 
$(D^* \bar D)_+^0$ channel were derived in Ref.~\cite{Braaten:2007dw}. 
We give here a more explicit derivation
of the line shapes produced by the decay $B^+ \to K^+ + X$. 
Our starting point is the optical theorem for the width of the $B^+$:
\begin{equation}
\Gamma [B^+] = 
- \frac 1 {M_B} \, {\rm Im} \, {\mathcal A}[ B^+ \to B^+] ,
\label{GamB:optical}
\end{equation}
where ${\mathcal A} [ B^+ \to B^+]$ is the one-meson-irreducible forward
amplitude for $B^+$. This amplitude has contributions from intermediate
states consisting of a $K^+$ recoiling against sets of particles whose
invariant mass $M_{*0} + M_0 + E$ is near the $D^{*0} \bar D^0$
threshold.  There is resonant enhancement for small $E$
if the particles are accessible
from the $(D^* \bar D)_+^0$ channel.  The resonant contributions
to the forward amplitude can be expressed as a loop integral over the
4-momentum $P_k$ of the $K^+$:
\begin{equation}
{\cal A}_{\rm res} [B^+ \to B^+] = 
- \int \frac {d^4 P_K}{(2 \pi)^4} 
\left( {\mathcal C}^{K^+}_{B^+} \, f(E) \, {\mathcal C}^{K^+}_{B^+} \right) 
\frac{i}{P_K^2 - m_K^2 + i \varepsilon}.
\label{Aforward}
\end{equation}
There is an implicit restriction of the integral to the region of small $E$.
The expression inside the parentheses takes into account
the amplitude for the creation of charm mesons in the channel 
$(D^* \bar D)^0_+$, the resonant propagation of the pair of charm mesons, 
and the amplitude for their annihilation. 
Factorization has been used to express it as the
product of a long-distance factor and two short-distance factors. The
long-distance factor $f(E)$ is the scattering amplitude for elastic
scattering in the $(D^* \bar D)_+^0$ channel given in Eq.~(\ref{f-E}).
The short-distance factors $C^{K^+}_{B^+}$ depend on the 4-momenta $P_B$
and $P_K$ of the $B^+$ and $K^+$, but they are insensitive to the small 
energy $E$ defined by $(P_B - P_K)^2 = (M_{*0} + M_0 + E)^2$.
The short-distance factors can therefore be simplified by setting $E=0$.  
Using the Cutkosky
cutting rules, the resonant contribution to the imaginary part of the
forward amplitude can be written
\begin{equation}
{\rm Im} {\mathcal A}_{\rm res} [ B^+ \to B^+ ] = 
- \int \frac {d^3 P_K}{(2 \pi)^3 2 E_K} \, 
\left( {\mathcal C}^{K^+}_{B^+} \, {\rm Im} \, f(E) \, ({\mathcal C}^{K^+}_{B^+})^* \right) .
\label{ImAforward}
\end{equation}
Again there is an implied restriction of the integral to the region of
small $E$. The contribution to the width of $B^+$ from its decay into
$K^+$ and the $X(3872)$ resonance can be obtained by
inserting Eq.~(\ref{ImAforward}) into Eq.~(\ref{GamB:optical}). 
The distribution in the invariant mass $M=M_{*0} + M_0 + E$ 
of the resonance can be obtained by inserting the
identity
\begin{equation}
1 = \int {d^4 P_R} \, \delta^4 (P_B - P_K - P_R) 
	\int d M^2 \, \delta (M^2 - P_R^2 ) .
\end{equation}
Changing the order of integration and using $E_B - E_K >0$ and 
$|E| \ll M_{*0} + M_0$, this can be written
\begin{equation}
1 = \frac {M_{*0} + M_0}{\pi} \int d E 
\int \frac {d^3 P_R}{(2 \pi)^3 2E_R} (2 \pi)^4 \delta^4 (P_B - P_K - P_R) .
\end{equation}
Upon inserting this into Eq.~(\ref{ImAforward}),
we obtain a factorization formula for the inclusive energy
distribution summed over all resonant channels:
\begin{equation}
\frac {d \Gamma}{dE} [ B^+ \to K^+ + {\rm resonant} ] = 2 \,
\Gamma_{B^+}^{K^+} \, {\rm Im} \, f(E) .
\label{lsh1:res}
\end{equation}
The short-distance factor is a positive real constant:
\begin{equation}
\Gamma_{B^+}^{K^+} = \frac {M_{*0} + M_0} {2 \pi M_B} \int \frac {d^3
P_k} {(2 \pi)^3 2E_R} \int \frac {d^3 P_K}{(2 \pi)^3 2E_R} (2 \pi)^4
\delta^4 (P_B - P_k - P_R) |{\mathcal C}^{K^+}_{B^+} |^2 .
\end{equation}

A more explicit expression for the short-distance factor
can be obtained by using Lorentz invariance to express 
the short-distance factor ${\mathcal C}^{K^+}_{B^+}$ in the form 
\begin{equation}
{\mathcal C}^{K^+}_{B^+} = C^{K^+}_{B^+} P_B \cdot (\epsilon_{D^*})^* ,
\end{equation}
where $\epsilon_{D^*}$ is a polarization vector for the $D^{*0}$ 
\cite{Braaten:2004fk,Braaten:2004ai} and $C^{K^+}_{B^+}$ 
is a constant with dimensions of inverse mass. 
Evaluating the phase space integral and summing over the 
$D^{*0}$ spins, we get
\begin{equation}
\Gamma^{K^+}_{B^+} = 
\frac {\lambda^{3/2} (M_B, m_K, M_{*0}+M_0)}
{64 \pi^2 (M_{*0} +M_0) M^3_B} \left| C^{K^+}_{B^+} \right|^2 .
\end{equation}

The optical theorem in Eq.~(\ref{ImA-optical2}) can be used to
resolve the inclusive resonant rate in Eq.~(\ref{lsh1:res}) 
into two terms according to whether they have 
${\rm Im} \gamma$ or ${\rm Im} \kappa(E)$ as a factor. 
We interpret the term proportional to ${\rm Im} \gamma$
as the contribution from all short-distance decay 
channels $C$.  The imaginary part of $\gamma$ can be 
expressed as a sum over those decay channels:
\begin{equation}
{\rm Im} \gamma = \sum_C \Gamma^C(E).
\label{Imgam-Gam}
\end{equation}
We have allowed for the possibility that the dependence of 
some of the short-distance factors $\Gamma^C(E)$
on the energy $E$ may not be negligible in the $D^{*0} \bar D^0$ 
threshold region.
Thus the energy distribution in a specific short-distance 
channel $C$ can be expressed as
\begin{equation}
\frac{d\Gamma}{dE}[ B^+ \to K^+ + C] =
2 \, \Gamma_{B^+}^{K^+} \, |f(E)|^2 \, \Gamma^C(E).
\label{lsh1:C}
\end{equation}
We interpret the term in Eq.~(\ref{lsh1:res}) proportional 
to ${\rm Im} \kappa(E)$
as the contribution from channels that correspond to 
$D^{*0} \bar D^0$ or $D^0 \bar D^{*0}$ followed by the decay of 
the $D^{*0}$ or $\bar D^{*0}$.  We can resolve this 
term into the contributions from the channels 
$D^0 \bar D^0 \pi^0$, $D^+ \bar D^0 \pi^-$,
$D^0 D^- \pi^+$, and $D^0 \bar D^0 \gamma$ by multiplying it by the
energy-dependent branching fractions ${\rm Br}_{000}(E)$, 
$\frac{1}{2} {\rm Br}_{011}(E)$, $\frac{1}{2} {\rm Br}_{011}(E)$, 
and ${\rm Br}_{00\gamma}(E)$, which add up to 1.
A simple expression for ${\rm Im} \kappa(E)$ can be obtained 
by using the identity in Eq.~(\ref{sqrtid}).
The resulting expression for the energy distribution in the 
$D^0 \bar D^0 \pi^0$ channel is 
\begin{equation}
\frac{d\Gamma}{dE}[ B^+ \to K^+ + D^0 \bar D^0 \pi^0] =
2 \, \Gamma_{B^+}^{K^+} \, | f(E) |^2
\left[ M_{*00} \big( \sqrt{E^2 + \Gamma_{*0}(E)^2/4} + E \big) \right]^{1/2}
{\rm Br}_{000}(E) ,
\label{lsh1:DDbarpi}
\end{equation}
where ${\rm Br}_{000}(E)$ is given in Eq.~(\ref{Br000}).

The energy distributions of the $X$ resonance
in the decays $B^0 \to K^0 + X$ 
are given by expressions identical to those in 
Eqs.~(\ref{lsh1:res}), (\ref{lsh1:C}), and 
(\ref{lsh1:DDbarpi}) except that the short-distance constant
$\Gamma_{B^+}^{K^+}$ is replaced by $\Gamma_{B^0}^{K^0}$.
Thus the line shapes for $X(3872)$ produced in $B^+$ decays 
and $B^0$ decays are predicted to be identical in the region
within a few MeV of the $D^{*0} \bar D^0$ threshold.

In Ref.~\cite{Braaten:2005ai}, the decay rates of $X$ into $J/\psi$
plus $\pi^+ \pi^-$, $\pi^+ \pi^- \pi^0$, $\pi^0 \gamma$, and 
$\gamma$ were calculated under the assumption that
these decays proceed through couplings of the $X$ to $J/\psi$
and the vector mesons $\rho^0$ and $\omega$.
The results of Ref.~\cite{Braaten:2005ai} were used in 
Ref.~\cite{Braaten:2007dw} to calculate
the dependence of the factor $\Gamma^{C}(E)$ 
in Eq.~(\ref{lsh1:C}) on the energy $E$
for $C= J/\psi \, \pi^+ \pi^-$ and $J/\psi \, \pi^+ \pi^- \pi^0$.
The normalization factors $\Gamma^{C}(0)$ can only be determined 
by measurements of $X(3872)$ decays. 
Simple analytic approximations to $\Gamma^{C}(E)/\Gamma^{C}(0)$
that are accurate to within 1\% in the region $|E|< 8.5$ MeV  for
$J/\psi \, \pi^+ \pi^-$ and in the region $|E| < 1$ MeV for
$J/\psi \, \pi^+ \pi^- \pi^0$ are given in Ref.~\cite{Braaten:2007dw}.

\begin{figure}[t]
\includegraphics[width=12cm,clip=true]{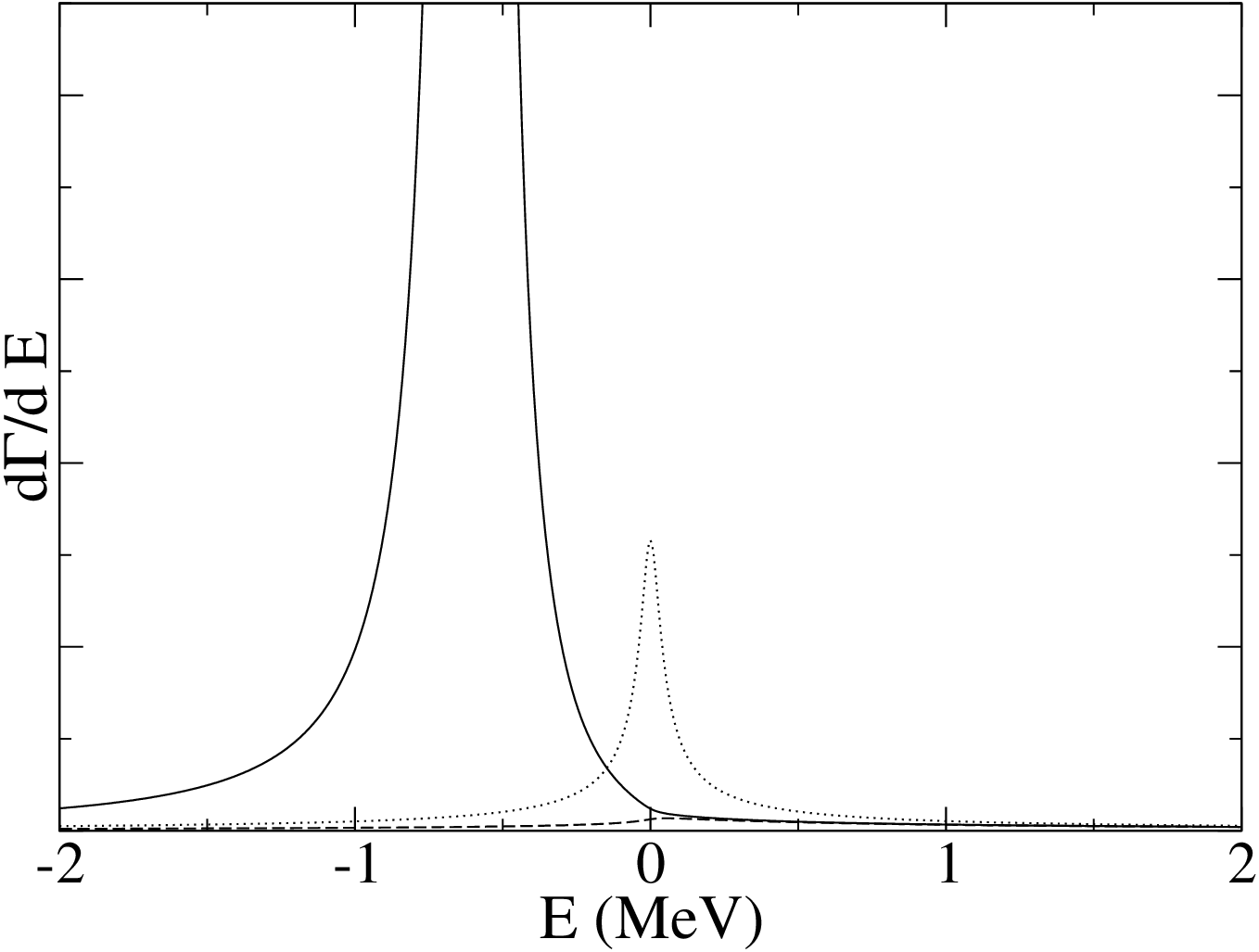}
\caption{The line shapes near the $D^{*0} \bar D^0$ threshold
for $X(3872)$ decaying into 
a short-distance channel, such as $J/\psi \, \pi^+ \pi^-$ 
or $J/\psi \, \pi^+ \pi^- \pi^0$.  
The line shapes are shown for three values of $\gamma$:
+34 MeV (solid line), 0 (dotted line), and $-34$ MeV (dashed line).
\label{fig:LSsd} }
\end{figure}

\begin{figure}[t]
\includegraphics[width=12cm,clip=true]{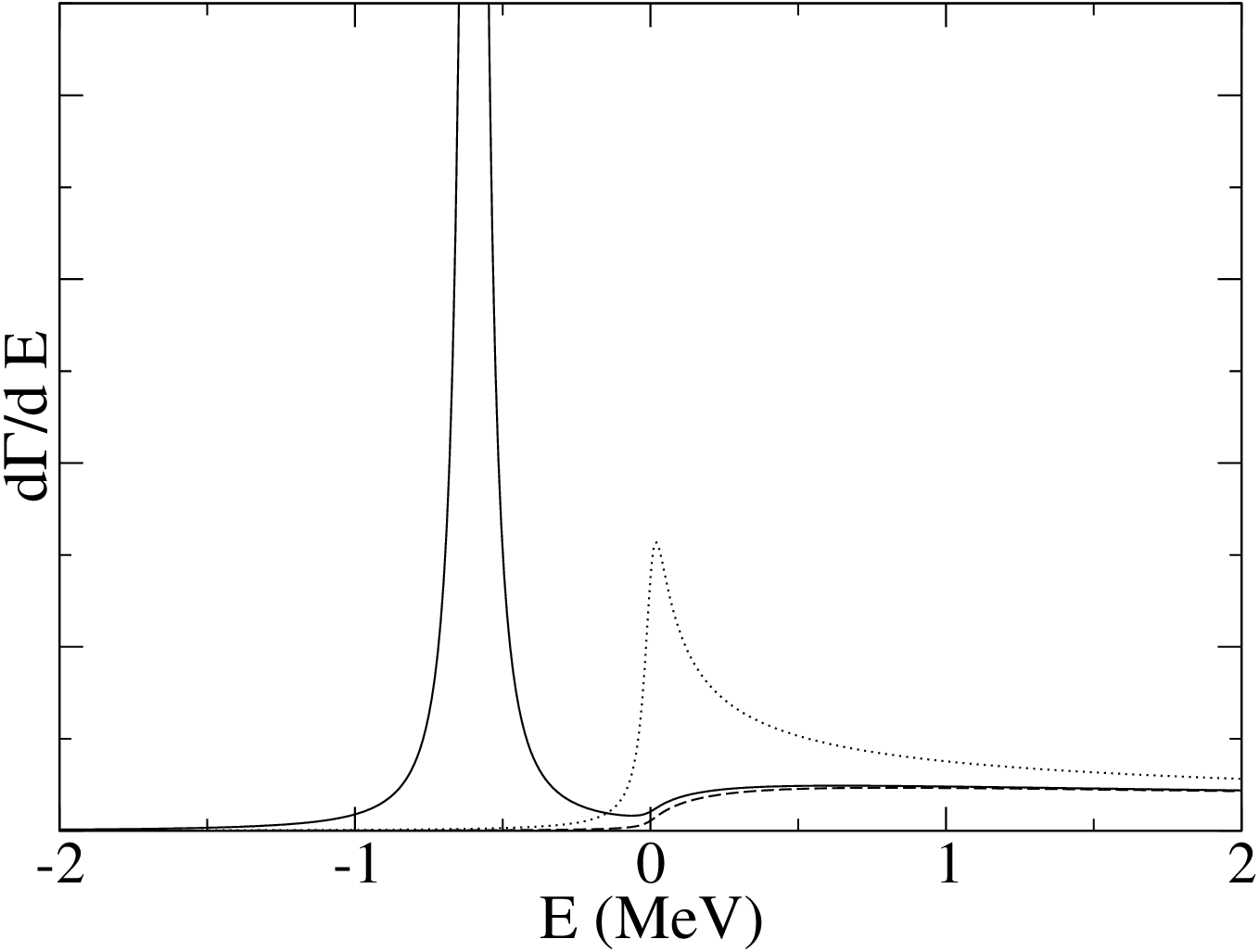}
\caption{The line shapes near the $D^{*0} \bar D^0$ threshold 
for $X(3872)$ in the $D^0 \bar D^0 \pi^0$ channel.  
The line shapes are shown for three values of $\gamma$:
+34 MeV (solid line), 0 (dotted line), and $-34$ MeV (dashed line).
\label{fig:LSld} }
\end{figure}

In Figs.~\ref{fig:LSsd} and \ref{fig:LSld}, we illustrate the line shapes 
for $X(3872)$  near the $D^{*0} \bar D^0$ threshold.
We take into account the $D^{*0}$ width, but we neglect the effect 
on the line shapes of inelastic scattering channels for the charm mesons.
We show the line shapes for three values of $\gamma$:
+34, 0, and $-34$ MeV.
For $\gamma = +34$ MeV, the peak of the resonance is at $E= -0.6$ MeV, 
which is the central value 
of the measurement in Eq.~(\ref{MX-CLEO}).
In Fig.~\ref{fig:LSsd}, we show the line shapes in a short-distance 
decay mode,  such as $J/\psi \, \pi^+ \pi^- \pi^0$ 
or $J/\psi \, \pi^+ \pi^-$.  The line shape is given by
Eq.~(\ref{lsh1:C}).  We have neglected the energy-dependence 
of the factor $\Gamma^C(E)$. 
The relative normalizations of the curves for the three values 
of $\gamma$ are determined by using the same short-distance factors 
$\Gamma_{B^+}^{K^+}$ and $\Gamma^C$.  For $\gamma = +34$ MeV,
which corresponds to a bound state, the line shape is dominated 
by the Breit-Wigner resonance near $E = -0.6$ MeV. For $\gamma = -34$ MeV,
which corresponds to a virtual state, the line shape has a cusp 
near $E = 0$ MeV.
In Fig.~\ref{fig:LSld}, we show the line shapes in 
the $D^0 \bar D^0 \pi^0$ channel.  The line shape is given by
Eq.~(\ref{lsh1:DDbarpi}). 
The relative normalizations of the curves for the three values 
of $\gamma$ are determined by using the same short-distance factor 
$\Gamma_{B^+}^{K^+}$.  For $\gamma = +34$ MeV,
which corresponds to a bound state, the dominant features of the
line shape are
a Breit-Wigner resonance near $E = -0.6$ MeV and a threshold 
enhancement for $E>0$. For $\gamma = -34$ MeV,
which corresponds to a virtual state, the line shape has 
only the threshold enhancement.

\subsection{Coupled neutral and charged channels}
\label{sec:lineshape2}

We proceed to generalize the factorization formulas
in Section~\ref{sec:lineshape1} to the two-channel case.
We begin by generalizing the forward amplitude in 
Eq.~(\ref{Aforward}).  We have to take into account the
possibility of resonant scattering between any pair of the 
charged and neutral channels.  The amplitude can be written
\begin{equation}
{\cal A}_{\rm res} [B^+ \to B^+] = 
- \int \frac {d^4 P_K}{(2 \pi)^4} \sum_{i=0}^1 \sum_{j=0}^1 
\left( {\mathcal C}^{K^+,i}_{B^+} \, f_{ij}(E) \, {\mathcal C}^{K^+,j}_{B^+} \right) 
\frac{i}{P_K^2 - m_K^2 + i \varepsilon}.
\label{Aforward2}
\end{equation}
Following the same path as in Section~\ref{sec:lineshape1},
we ultimately arrive at a factorization formula for the inclusive 
energy distribution summed over all resonant channels:
\begin{equation}
\frac{d\Gamma}{dE}[ B^+ \to K^+ + {\rm resonant}] =
2 \, \sum_{i=0}^1 \sum_{j=0}^1 \Gamma_{B^+}^{K^+,ij} \, {\rm Im} \, f_{ij}(E) .
\label{lsh2:res}
\end{equation}
The short-distance factors are
\begin{equation}
\Gamma_{B^+}^{K^+,ij} = 
\frac{M_{*0} + M_0}{2 \pi M_B} 
\int \frac{d^3 P_R}{(2 \pi)^3 2E_R} 
\int \frac{d^3 P_K}{(2 \pi)^3 2E_K} 
(2 \pi)^4 \delta^4 (P_B - P_K - P_R) \, 
{\mathcal C}^{K^+,i}_{B^+} \, ({\mathcal C}^{K^+,j}_{B^+})^* .
\end{equation}
The short-distance factors $\Gamma_{B^+}^{K^+,00}$
and $\Gamma_{B^+}^{K^+,11}$ are positive real constants, while
$\Gamma_{B^+}^{K^+,01} = (\Gamma_{B^+}^{K^+,10})^*$
is a complex constant.  Thus there are four independent real constants 
associated with the $B^+ \to K^+$ transitions.
These constants satisfy the Schwarz inequality
\begin{equation}
\left| \Gamma_{B^+}^{K^+,01} \right|^2 
\le \Gamma_{B^+}^{K^+,00} \, \Gamma_{B^+}^{K^+,11} .
\label{Schwartz}
\end{equation}

The optical theorem in Eq.~(\ref{Imfij-I})
can be used to resolve the inclusive resonant rate 
in  Eq.~(\ref{lsh2:res}) into four terms
according to whether they have ${\rm Im} \gamma_0$, 
${\rm Im} \gamma_1$, ${\rm Im} \kappa(E)$, 
or ${\rm Im} \kappa_1(E)$ as a factor. 
We interpret the terms proportional to ${\rm Im} \gamma_0$ 
and ${\rm Im} \gamma_1$ as the contributions 
from short-distance decay channels $C$.  The imaginary parts of 
$\gamma_0$ and $\gamma_1$ can be expressed as sums over those 
decay channels:
\begin{subequations}
\begin{eqnarray}
{\rm Im} \gamma_0 = \sum_C \Gamma_0^C(E),
\label{Imgam0-Gam}
\\
{\rm Im} \gamma_1 = \sum_C \Gamma_1^C(E).
\label{Imgam1-Gam}
\end{eqnarray}
\label{Imgam01-Gam}
\end{subequations}
The factorization formula for the energy distribution in a 
specific short-distance decay channel $C$ is
\begin{eqnarray}
\frac{d\Gamma}{dE}[ B^+ \to K^+ + C] &=&
\bigg( \sum_{i=0}^1 \sum_{j=0}^1 \, \Gamma_{B^+}^{K^+,ij} 
[f_{i0}(E) - f_{i1}(E)] [f_{j0}^*(E) - f_{j1}^*(E)] \bigg) \, \Gamma_0^C(E)
\nonumber
\\
&+& 
\left( \sum_{i=0}^1 \sum_{j=0}^1 \, \Gamma_{B^+}^{K^+,ij}
 [f_{i0}(E) + f_{i1}(E)] [f_{j0}^*(E) + f_{j1}^*(E)] \right)  \Gamma_1^C(E).
\label{lsh2:C}
\end{eqnarray}

The terms in Eq.~(\ref{lsh2:res}) proportional to ${\rm Im} \kappa(E)$
and ${\rm Im} \kappa_1(E)$ also have simple interpretations.
We interpret the term proportional to ${\rm Im} \kappa(E)$
as the contribution from channels that correspond to 
$D^{*0} \bar D^0$ or $D^0 \bar D^{*0}$ followed by the decay of 
the $D^{*0}$ or $\bar D^{*0}$.  We can resolve this 
term into the contributions from the individual channels 
$D^0 \bar D^0 \pi^0$, $D^+ \bar D^0 \pi^-$, 
$D^0 D^- \pi^+$, and $D^0 \bar D^0 \gamma$ by multiplying it by the
energy-dependent branching fractions 
${\rm Br}_{000}(E)$, $\frac{1}{2} {\rm Br}_{011}(E)$, and
$\frac{1}{2} {\rm Br}_{011}(E)$, and ${\rm Br}_{00\gamma}(E)$,
which add up to 1.  For example, the line shape of $X$ in the 
$D^0 \bar D^0 \pi^0$ decay mode is 
\begin{eqnarray}
\frac{d\Gamma}{dE}[ B^+ \to K^+ + D^0 \bar D^0 \pi^0] &=&
2 \bigg( \sum_{i=0}^1 \sum_{j=0}^1 \, \Gamma_{B^+}^{K^+,ij}
	f_{i0}(E) f_{j0}^*(E) \bigg) 
\nonumber
\\
&& \times 
\left[ M_{*00} \big( \sqrt{E^2 + (\Gamma_{*0}(E)/2)^2} + E \big) \right]^{1/2}
{\rm Br}_{000}(E) ,
\label{lsh2:DDbarpi}
\end{eqnarray}
where ${\rm Br}_{000}(E)$ is given in Eq.~(\ref{Br000}).
We interpret the term in Eq.~(\ref{lsh2:res}) 
proportional to ${\rm Im} \kappa_1(E)$
as the contribution from channels that correspond to 
$D^{*+} D^-$ or $D^+ D^{*-}$ followed by the decay of 
the $D^{*+}$ or $D^{*-}$.  We can resolve this 
term into the contributions from the individual channels $D^+ D^- \pi^0$,
$D^0 D^- \pi^+$, $D^+ \bar D^0 \pi^-$, 
and $D^+ D^- \gamma$ by multiplying it by
energy-dependent branching fractions.  For example, the line shape 
of $X$ in the $D^+ D^- \pi^0$ decay mode is 
\begin{eqnarray}
\frac{d\Gamma}{dE}[ B^+ \to K^+ + D^+ D^- \pi^0] &=&
2 \bigg( \sum_{i=0}^1 \sum_{j=0}^1 \, \Gamma_{B^+}^{K^+,ij} 
	f_{i1}(E) f_{j1}^*(E) \bigg)
\nonumber
\\
&& \hspace{-2cm} \times 
\left[ M_{*11} \big( \sqrt{(E - \nu)^2 + (\Gamma_{*1}(E - \nu)/2)^2} 
	+ E - \nu \big) \right]^{1/2}
{\rm Br}_{110}(E) ,
\label{lsh2:DpDmpi}
\end{eqnarray}
where ${\rm Br}_{110}(E)$ is given in Eq.~(\ref{Br110}).
The expressions for the line shapes in the decay channels 
$D^0 D^- \pi^+$ and $D^- D^0 \pi^+$ are more complicated
because they receive contributions from channels that correspond 
to $D^{*0} \bar D^0$ or $D^0 \bar D^{*0}$ as well as channels 
that correspond to $D^{*+} D^-$ or $D^+ D^{*-}$. 

If the energy $E$ is very close to the $D^{*0} \bar D^0$ threshold, 
the two-channel factorization formulas in Eqs.~(\ref{lsh2:res}), 
(\ref{lsh2:C}), and (\ref{lsh2:DDbarpi}) should reduce to
the single-channel factorization formulas in Eqs.~(\ref{lsh1:res}),
(\ref{lsh1:C}), and (\ref{lsh1:DDbarpi}).   For the factorization formula
for $B^+ \to K^+ + D^0 \bar D^0 \pi^0$ in Eq.~(\ref{lsh2:DDbarpi}),
this can be verified by inserting the expressions in 
Eq.~(\ref{fij:factor}) for the scattering amplitudes $f_{ij}(E)$ 
at small $E$.  The factorization formula
reduces to Eq.~(\ref{lsh1:DDbarpi})
with the short-distance factor $\Gamma _{B^+}^{K^+}$ given by
\begin{equation}
\Gamma _{B^+}^{K^+} \approx 
\sum_{i=0}^1 \sum_{j=0}^1 \, \Gamma _{B^+}^{K^+, ij} c_i c_j^* .
\label{GamBK-red}
\end{equation}
Similarly, the factorization formula for $B^+ \to K^+ + C$ in 
Eq.~(\ref{lsh2:C}) reduces to Eq.~(\ref{lsh1:C}) with 
$\Gamma _{B^+}^{K^+}$ given by  
Eq.~(\ref{GamBK-red}) and $\Gamma^C(E)$ given by
\begin{equation}
\Gamma^C (E) \approx \frac{|1-c_1|^2}{2} \Gamma_0^C (E) 
+ \frac{|1+c_1|^2}{2} \Gamma_1^C (E) .
\end{equation}
To see that the two-channel factorization formula in Eq.~(\ref{lsh2:res}) 
for the inclusive resonant rate reduces to Eq.~(\ref{lsh1:res}),  
we express the imaginary part of $f_{ij}(E)$ 
in a form that is compatible with the Cutkosky cutting rules:
\begin{equation}
{\rm Im} f_{ij}(E) = 
c_i \, f(E) \left( {\rm Im} c_j \right) 
+ c_i \left( {\rm Im} f(E) \right) c^*_j 
+ \left( {\rm Im} c_i \right) f^*(E) \, c^*_j .
\label{Imfij:cut}
\end{equation}
Since $c_0 = 1$, it has no imaginary part. The expression for
$c_1$ in Eq.~(\ref{c1-factor}) is a function of $\gamma_1$ and 
$\kappa_1 (0)$ only. The imaginary part of $\kappa_1(0)$ is 
suppressed relative to its real part by a factor of 
$\Gamma[D^{*+}]/\nu$.  We expect $\gamma_1$ to have a real part 
that is comparable to or larger than $m_\pi$, so the 
imaginary part of $\gamma_1$ should also be small relative 
to its real part. Thus the only term on the right side of 
Eq.~(\ref{Imfij:cut}) that is not suppressed is the one with 
the factor ${\rm Im}f(E)$.  Inserting that term into the 
factorization formula in Eq.~(\ref{lsh2:res}), 
we find that it reduces to Eq.~(\ref{lsh1:res}) with 
$\Gamma_{B^+}^{K^+}$ given by Eq.~(\ref{GamBK-red}).

The factorization formulas for the energy distributions simplify 
if the parameter $\gamma_1$ is assumed to be large compared to 
$\kappa_1(0)$.  The scattering amplitudes $f_{ij}(E)$ in Eq.~(\ref{fij-E})
reduce to the expressions in Eq.~(\ref{fij:largegam1}).
The two-channel
factorization formula in Eqs.~(\ref{lsh2:res}), (\ref{lsh2:C}), and 
(\ref{lsh2:DDbarpi}) all reduce to the single-channel factorization 
formulas in Eqs.~(\ref{lsh1:res}), (\ref{lsh1:C}), 
and Eq.~(\ref{lsh1:DDbarpi}) with the scattering amplitude $f(E)$
replaced by the expression for $f_{00}(E)$ given in 
Eq.~(\ref{fij:largegam1}).  By using Eq.~(\ref{gamma0-approx2})
to eliminate $\gamma_0$ in favor of $\gamma$, 
the scattering amplitude reduces to 
\begin{equation}
f(E) \approx \frac{1}{-\gamma + \kappa(E) + \kappa_1(E) - \kappa_1(0)} .
\end{equation}
The short-distance factor for $B^+ \to K^+$ transitions reduces to
\begin{equation}
\Gamma_{B^+}^{K^+} \approx 
\sum_{i=0}^1 \sum_{J=0}^1 (-1)^{i+j} \Gamma_{B^+}^{K^+,ij} .
\end{equation}
The sums project the $(D^*\bar D)_+^i$ channels onto isospin 0.
The short-distance factor for the short-distance decay channel 
reduces to 
\begin{equation}
\Gamma^C(E) \approx 2 \Gamma^C_0(E) .
\end{equation}
The coefficient of $\Gamma^C_1(E)$ goes to zero in this limit.
Thus the decay of $X$ into final states $C$ with total isospin 
quantum number $I=1$, such as $J/\psi \, \pi^+ \pi^-$,
are suppressed in the large-$\gamma_1$ limit.

\subsection{Constraints from isospin symmetry}
\label{sec:isospin}

We have not yet fully exploited the approximate isospin symmetry of QCD.
Since the short-distance factors only involve momenta of order 
$m_\pi$ and larger, isospin-violating effects can be neglected
in these factors.  Thus isospin symmetry can be used 
to constrain the short-distance factors.
At the quark level, the transitions $B \to K + D^* \bar D$
and $B \to K + D \bar D^*$ proceed through two operators in the 
effective weak Hamiltonian:  the charged current operator
$\bar b \gamma^\mu (1 - \gamma_5) c \, \bar c \gamma_\mu (1 - \gamma_5) s$
and the neutral current operator
$\bar b \gamma^\mu (1 - \gamma_5) s \, \bar c \gamma_\mu (1 - \gamma_5) c$.
These operators are both isospin singlets.
Thus isospin symmetry is respected by these transitions.
It can therefore be used to relate the 
short-distance coefficients $\mathcal{C}_{B^+}^{K^+,i}$
for the $B^+ \to K^+$ transition to the 
short-distance coefficients $\mathcal{C}_{B^0}^{K^0,i}$
for the $B^0 \to K^0$ transition.
Since $B^+$ and $B^0$ form an isospin doublet and 
$K^+$ and $K^0$ form an isospin doublet, the coefficients
$\mathcal{C}_{B^0}^{K^0,i}$ and $\mathcal{C}_{B^+}^{K^+,i}$ 
are related by Clebsch-Gordan coefficients:
\begin{subequations}
\begin{eqnarray}
\mathcal{C}_{B^0}^{K^0,0} &=& - \mathcal{C}_{B^+}^{K^+,1},
\label{C:BK0}
\\
\mathcal{C}_{B^0}^{K^0,1} &=& - \mathcal{C}_{B^+}^{K^+,0}.
\label{C:BK1}
\end{eqnarray}
\label{CBK:isospin}
\end{subequations}
This implies that the short-distance constants
$\Gamma_{B^0}^{K^0,ij}$ in the factorization formulas 
for $B^0 \to K^0$ transitions 
are related to the corresponding constants
$\Gamma_{B^+}^{K^+,ij}$ in the factorization formulas 
for $B^+ \to K^+$ transitions by
\begin{subequations}
\begin{eqnarray}
\Gamma_{B^0}^{K^0,00} &=& \Gamma_{B^+}^{K^+,11},
\\
\Gamma_{B^0}^{K^0,01} &=& ( \Gamma_{B^+}^{K^+,01} )^*,
\\
\Gamma_{B^0}^{K^0,11} &=& \Gamma_{B^+}^{K^+,00} .
\end{eqnarray}
\label{GamBK:isospin}
\end{subequations}
Thus the short-distance constants associated with 
the $B^+ \to K^+$ and $B^0 \to K^0$ transitions are determined
by four independent real constants.

Isospin symmetry also constrains the short-distance factors
$\Gamma_I^C(E)$ associated with decays of $X$ into short-distance 
decay modes.  It implies that
for a decay channel $C$ with definite isospin 
quantum number $I=0$ or $I=1$,
only the term with the factor $\Gamma_I(E)$ contributes.
An example of a decay channel 
with isospin quantum number $I=0$ is $J/\psi \, \pi^+ \pi^- \pi^0$, 
assuming that the $\pi^+ \pi^- \pi^0$ 
comes from the decay of a virtual $\omega$.  
An example of a decay channel
with isospin quantum number $I=1$ is $J/\psi \, \pi^+ \pi^-$, 
assuming that the $\pi^+ \pi^-$ 
comes from the decay of a virtual $\rho^0$.
We will give the factorization formulas for short-distance 
decay channels with definite isospin quantum number
$I=0$ and $I=1$ for both $B^+ \to K^+$ transitions
and $B^0 \to K^0$ transitions.
For a short-distance decay channel $C$ with definite isospin quantum number
$I=0$, such as $J/\psi \, \pi^+ \pi^- \pi^0$, 
the energy distribution in Eq.~(\ref{lsh2:C}) reduces to
\begin{subequations}
\begin{eqnarray}
\frac{d\Gamma}{dE}[ B^+ \to K^+ + C] &=&  
4\left( \Gamma_{B^+}^{K^+,00} |\gamma_1 - \kappa_1(E)|^2 
-  2 {\rm Re} [\Gamma_{B^+}^{K^+,01} (\gamma_1 - \kappa_1(E)) (\gamma_1 - \kappa(E))^*]
\right.
\nonumber
\\ 
&& \hspace{4cm} \left.
+ \Gamma_{B^+}^{K^+,11} |\gamma_1 - \kappa(E)|^2 \right) 
\frac{\Gamma^C_0(E)}{|D(E)|^2} ,
\label{lsh:B+I=0}
\\
\frac{d\Gamma}{dE}[ B^0 \to K^0 + C] &=&  
4 \left( \Gamma_{B^+}^{K^+,00} |\gamma_1 - \kappa(E)|^2 
- 2 {\rm Re} [\Gamma_{B^+}^{K^+,01} (\gamma_1 - \kappa(E)) (\gamma_1 - \kappa_1(E))^*]
\right.
\nonumber
\\ 
&& \hspace{4cm} \left.
+ \Gamma_{B^+}^{K^+,11} |\gamma_1 - \kappa_1(E)|^2 \right) 
\frac{\Gamma^C_0(E)}{|D(E)|^2} .
\label{lsh:B0I=0}
\end{eqnarray}
\label{lsh:BI=0}
\end{subequations}
For a short-distance decay channel $C$ with definite isospin quantum number
$I=1$, such as $J/\psi \, \pi^+ \pi^-$,  
the energy distribution in Eq.~(\ref{lsh2:C}) reduces to
\begin{subequations}
\begin{eqnarray}
\frac{d\Gamma}{dE}[ B^+ \to K^+ + C] &=&  
4 \left( \Gamma_{B^+}^{K^+,00} |\gamma_0 - \kappa_1(E)|^2 
+ 2 {\rm Re} [\Gamma_{B^+}^{K^+,01} (\gamma_0 - \kappa_1(E)) (\gamma_0 - \kappa(E))^*]
\right.
\nonumber
\\ 
&& \hspace{4cm} \left.
+ \Gamma_{B^+}^{K^+,11} |\gamma_0 - \kappa(E)|^2 \right) 
\frac{\Gamma^C_1(E)}{|D(E)|^2} ,
\label{lsh:B+I=1}
\\
\frac{d\Gamma}{dE}[ B^0 \to K^0 + C] &=&  
4 \left( \Gamma_{B^+}^{K^+,00} |\gamma_0 - \kappa(E)|^2 
+ 2 {\rm Re} [\Gamma_{B^+}^{K^+,01} (\gamma_0 - \kappa(E)) (\gamma_0 - \kappa_1(E))^*]
\right.
\nonumber
\\ 
&& \hspace{4cm} \left.
+ \Gamma_{B^+}^{K^+,11} |\gamma_0 - \kappa_1(E)|^2 \right) 
\frac{\Gamma^C_1(E)}{|D(E)|^2} .
\label{lsh:B0I=1}
\end{eqnarray}
\label{lsh:BI=1}
\end{subequations}
In Eqs.~(\ref{lsh:B0I=0}) and (\ref{lsh:B0I=1}), we have used the isospin 
symmetry relations in Eqs.~(\ref{GamBK:isospin}) to express the 
short-distance coefficients $\Gamma_{B^0}^{K^0,ij}$ 
in terms of $\Gamma_{B^+}^{K^+,ij}$.

The effects of the charged charm meson channel 
$(D^* \bar D)_+^1$ on the line shapes of $X(3872)$ in the decays 
$B \to K + J/\psi \, \pi^+ \pi^-$ and 
$B \to K + J/\psi \, \pi^+ \pi^- \pi^0$ have been
discussed recently by Voloshin \cite{Voloshin:2007hh}. 
Voloshin made conceptual errors by ignoring resonant scattering 
between the $(D^* \bar D)_+^0$ and $(D^* \bar D)_+^1$ channels
and ignoring the constraints of isospin symmetry 
on the transitions $B \to K$.
In Voloshin's paper, our parameters $\gamma_0$ and $\gamma_1$ 
are denoted by $\kappa_0$ and $\kappa_1$ and the analogs of our functions
$\kappa(E)$ and $\kappa_1(E$) are denoted by $- i k_n$ and $\kappa_c$.
Voloshin took into account the constraints of isospin symmetry 
associated with the $J/ \psi \, \pi^+ \pi^-$ 
and $J/\psi \, \pi^+ \pi^- \pi^0$ in the final state.  
His results for the energy distributions can be expressed 
in the form
\begin{subequations}
\begin{eqnarray}
\frac{d\Gamma}{dE}[ B \to K + J/\psi \, \pi^+ \pi^- \pi^0] &=&   
\left| \frac{\gamma_1 - \kappa_1(E)}{D(E)} \right|^2  
\Phi^{J/\psi \, \omega} ,
\label{lshV:Bpsi3pi}
\\
\frac{d\Gamma}{dE}[ B \to K + J/\psi \, \pi^+ \pi^-] &=&   
\left| \frac{\gamma_0 - \kappa_1(E)}{D(E)} \right|^2 \, 
\Phi^{J/\psi \, \rho} ,
\label{lshV:Bpsi2pi}
\end{eqnarray}
\label{lshV:Bpsinpi}
\end{subequations}
where $\kappa(E)= (- 2 M_{*00} E - i \varepsilon)^{1/2}$ and 
$\kappa_1(E)  = (- 2 M_{*11} (E - \nu) - i \varepsilon)^{1/2}$.
The normalizing factors $\Phi^{J/\psi \, \rho}$ 
and $\Phi^{J/\psi \, \omega}$
can presumably be different for $B^+$ decays and $B^-$ decays, 
although this was not stated explicitly in Ref.~\cite{Voloshin:2007hh}.
The line shapes however were predicted to be the same
for $B^+$ decays and $B^-$ decays. 
Voloshin's results in Eqs.~(\ref{lshV:Bpsinpi}) correspond to specific 
choices for the short-distance factors $\Gamma_{B^+}^{K^+,ij}$ 
in our general factorization formulas in Eqs.~(\ref{lsh:BI=0})
and (\ref{lsh:BI=1}). 
In the case of $B^+$ decays, his results in Eqs.~(\ref{lshV:Bpsinpi})
are consistent with our factorization formulas in Eqs.~(\ref{lsh:B+I=0}) 
and (\ref{lsh:B+I=1}) if $\Gamma_{B^+}^{K^+,00}$ 
is the only nonzero short-distance factor for the $B \to K$ transition.
In the case of $B^0$ decays, his results in Eqs.~(\ref{lshV:Bpsinpi})
are consistent with our factorization 
formulas in Eqs.~(\ref{lsh:B0I=0}) and (\ref{lsh:B0I=1}) if 
$\Gamma_{B^+}^{K^+,11}= \Gamma_{B^0}^{K^0,00}$ 
is the only such nonzero factor.  However these conditions 
for $B^+$ decays and $B^0$ decays are inconsistent.
Thus Voloshin's results are incompatible with the constraints 
of isospin symmetry associated with  the $B \to K$ transitions.
The primary conceptual error in Ref.~\cite{Voloshin:2007hh}
was the assumption that there is a resonance in the amplitude 
only if the $B \to K$ transition creates the charm mesons in the 
neutral channel $(D^* \bar D)_+^0$.  However there is also a 
resonant contribution coming from the $B \to K$ transition creating
charm mesons in the charged channel $(D^* \bar D)_+^1$ 
followed by the resonant scattering of the charm mesons 
into the neutral channel.
A second conceptual error in Ref.~\cite{Voloshin:2007hh}
was the failure to take into account the constraints of isospin 
symmetry on the amplitudes for the  $B \to K$ transition.

\subsection{Current-current factorization and heavy-quark symmetry}

In Ref.~\cite{Braaten:2004ai}, it was pointed out that the 
combination of a standard current-current factorization approximation 
and heavy-quark symmetry could be used to simplify the 
factorization formulas
associated with the $X$ resonance in $B \to K$ transitions.  
In the standard current-current factorization approximation,
the matrix elements
of the relevant terms in the effective weak Hamiltonian
are expressed as products of matrix elements of currents:
\begin{subequations}
\begin{eqnarray}
\langle K D^* \bar D | \bar b \gamma^\mu (1 - \gamma_5) c \, 
	\bar c \gamma_\mu (1 - \gamma_5) s | B \rangle &\approx& 
\langle \bar D | \bar b \gamma^\mu (1 - \gamma_5) c  | B \rangle \, 
	\langle K D^* | \bar c \gamma_\mu (1 - \gamma_5) s | 0 \rangle ,
\nonumber
\\
\label{fact-charged}
\\
\langle K D^* \bar D | \bar b \gamma^\mu (1 - \gamma_5) s \, 
	\bar c \gamma_\mu (1 - \gamma_5) c | B \rangle &\approx& 
\langle K | \bar b \gamma^\mu (1 - \gamma_5) s | B \rangle  \, 
	\langle D^* \bar D | \bar c \gamma_\mu (1 - \gamma_5) c | 0 \rangle .
\nonumber
\\
\label{fact-neutral}
\end{eqnarray}
\label{fact}
\end{subequations}
The $D^*$ and $\bar D$ in the final state can equally well be 
replaced by $D$ and $\bar D^*$.
The matrix element of the charged current 
$\bar b \gamma^\mu (1 - \gamma_5) c$ in Eq.~(\ref{fact-charged})
is nonzero only if the $\bar D$ contains the same light quark 
as the $B$.  In the case of a $B^+$, the $\bar D$ or $\bar D^*$ 
must be $\bar D^0$ or $\bar D^{*0}$.  In the case of a $B^0$, 
the $\bar D$ or $\bar D^*$ must be $D^-$ or $D^{*-}$.
As pointed out in Ref.~\cite{Braaten:2004ai}, heavy-quark symmetry 
implies that the matrix element of the neutral current
$\bar c \gamma_\mu (1 - \gamma_5) c$ in Eq.~(\ref{fact-neutral})
vanishes at the $D^* \bar D$ threshold. Thus this matrix element 
is suppressed in the $D^* \bar D$ threshold region.
Putting these two observations together, we conclude that the 
current-current factorization approximation together with 
heavy quark symmetry puts strong constriants on the matrix 
elements of the effective weak Hamiltonian.  It implies that 
in $B^+ \to K^+$ transitions, the formation of the $X(3872)$ 
resonance is dominated by the creation of charm mesons at 
short distances in the neutral channel $(D^* \bar D)_+^0$.  
Similarly, in $B^0 \to K^0$ transitions,  the formation of the $X(3872)$ 
resonance is dominated by the creation of charm mesons at 
short distances in the charged channel $(D^* \bar D)_+^1$.
These statements imply that the short-distance coefficients 
${\cal C}_{B^+}^{K^+,1}= - {\cal C}_{B^0}^{K^0,0}$
are suppressed relative to 
${\cal C}_{B^+}^{K^+,0}= - {\cal C}_{B^0}^{K^0,1}$.
This suppression leads to a hierarchy in the short-distance 
factors associated with $B \to K$ transitions in the 
factorization formulas:
\begin{equation}
\Gamma_{B^+}^{K^+,11} \ll |\Gamma_{B^+}^{K^+,01}| \ll \Gamma_{B^+}^{K^+,00} .
\label{GamBK:hierarchy}
\end{equation}

If we assume that $\Gamma_{B^+}^{K^+,11}$ and $|\Gamma_{B^+}^{K^+,01}|$
are negligible compared to $\Gamma_{B^+}^{K^+,00}$,
the expressions for the line shapes of $X(3872)$ 
become rather simple.
For a short-distance decay channel $C$ with definite isospin 
quantum number $I=0$, such as $J/\psi \, \pi^+ \pi^- \pi^0$, 
the energy distributions in Eqs.~(\ref{lsh:BI=0}) reduce to
\begin{subequations}
\begin{eqnarray}
\frac{d\Gamma}{dE}[ B^+ \to K^+ + C] &\approx&   
4 \, \Gamma_{B^+}^{K^+,00} 
\left| \frac{\gamma_1 - \kappa_1(E)}{D(E)} \right|^2 \, 
\Gamma^C_0(E) ,
\label{lsh2f:B+psi3pi}
\\
\frac{d\Gamma}{dE}[ B^0 \to K^0 + C] &\approx&   
4 \, \Gamma_{B^+}^{K^+,00}
\left| \frac{\gamma_1 - \kappa(E)}{D(E)} \right|^2 \, 
\Gamma^C_0(E) .
\label{lsh2f:B0psi3pi}
\end{eqnarray}
\label{lsh2f:Bpsi3pi}
\end{subequations}
For a short-distance decay channel $C$ with definite isospin 
quantum number $I=1$, such as $J/\psi \, \pi^+ \pi^-$,  
the energy distributions in Eqs.~(\ref{lsh:BI=1}) reduce to
\begin{subequations}
\begin{eqnarray}
\frac{d\Gamma}{dE}[ B^+ \to K^+ + C] &\approx&   
4 \, \Gamma_{B^+}^{K^+,00} 
\left| \frac{\gamma_0 - \kappa_1(E)}{D(E)} \right|^2 \, 
\Gamma^C_1(E) ,
\label{lsh2f:B+psi2pi}
\\
\frac{d\Gamma}{dE}[ B^0 \to K^0 + C] &\approx&   
4 \, \Gamma_{B^+}^{K^+,00} 
\left| \frac{\gamma_0 - \kappa(E)}{D(E)} \right|^2  
\Gamma^C_1(E) .
\label{lsh2f:B0psi2pi}
\end{eqnarray}
\label{lsh2f:Bpsi2pi}
\end{subequations}
For the $D^0 \bar D^0 \pi^0$ channel, the energy distribution
in Eqs.~(\ref{lsh2:DDbarpi}) from the $B^+ \to K^+$ transition and
its analog from the $B^0 \to K^0$ transition reduce to
\begin{subequations}
\begin{eqnarray}
\frac{d\Gamma}{dE}[ B^+ \to K^+ + D^0 \bar D^0 \pi^0] &\approx&
2 \, \Gamma_{B^+}^{K^+,00}
\left| \frac{\gamma_1 + \gamma_0 - 2 \kappa_1(E)}{D(E)} \right|^2 
\nonumber
\\
&&  \times 
\left[ M_{*00} \big( \sqrt{E^2 + \Gamma_{*0}(E)^2/4} + E \big) \right]^{1/2}
{\rm Br}_{000}(E) ,
\label{lsh2f:B+DDpi}
\\
\frac{d\Gamma}{dE}[ B^0 \to K^0 + D^0 \bar D^0 \pi^0] &\approx&
2 \, \Gamma_{B^+}^{K^+,00}
\left| \frac{\gamma_1 - \gamma_0}{D(E)} \right|^2  
\nonumber
\\
&&  \times  
\left[ M_{*00} \big( \sqrt{E^2 + \Gamma_{*0}(E)^2/4} + E \big) \right]^{1/2} \,
{\rm Br}_{000}(E) .
\label{lsh2f:B0DDpi}
\end{eqnarray}
\label{lsh2f:BDDpi}
\end{subequations}
Note that the line shapes in Eqs.~(\ref{lsh2f:Bpsi3pi}),
(\ref{lsh2f:Bpsi2pi}), and (\ref{lsh2f:BDDpi}) are determined
by the parameters $\gamma_0$ and $\gamma_1$ or, equivalently, 
$\gamma$ and $\gamma_1$.  The relative normalizations of the rates from 
the $B^0 \to K^0$ transition and from the $B^+ \to K^+$ transition
are also determined by $\gamma$ and $\gamma_1$.

\begin{figure}[t]
\includegraphics[width=12cm,clip=true]{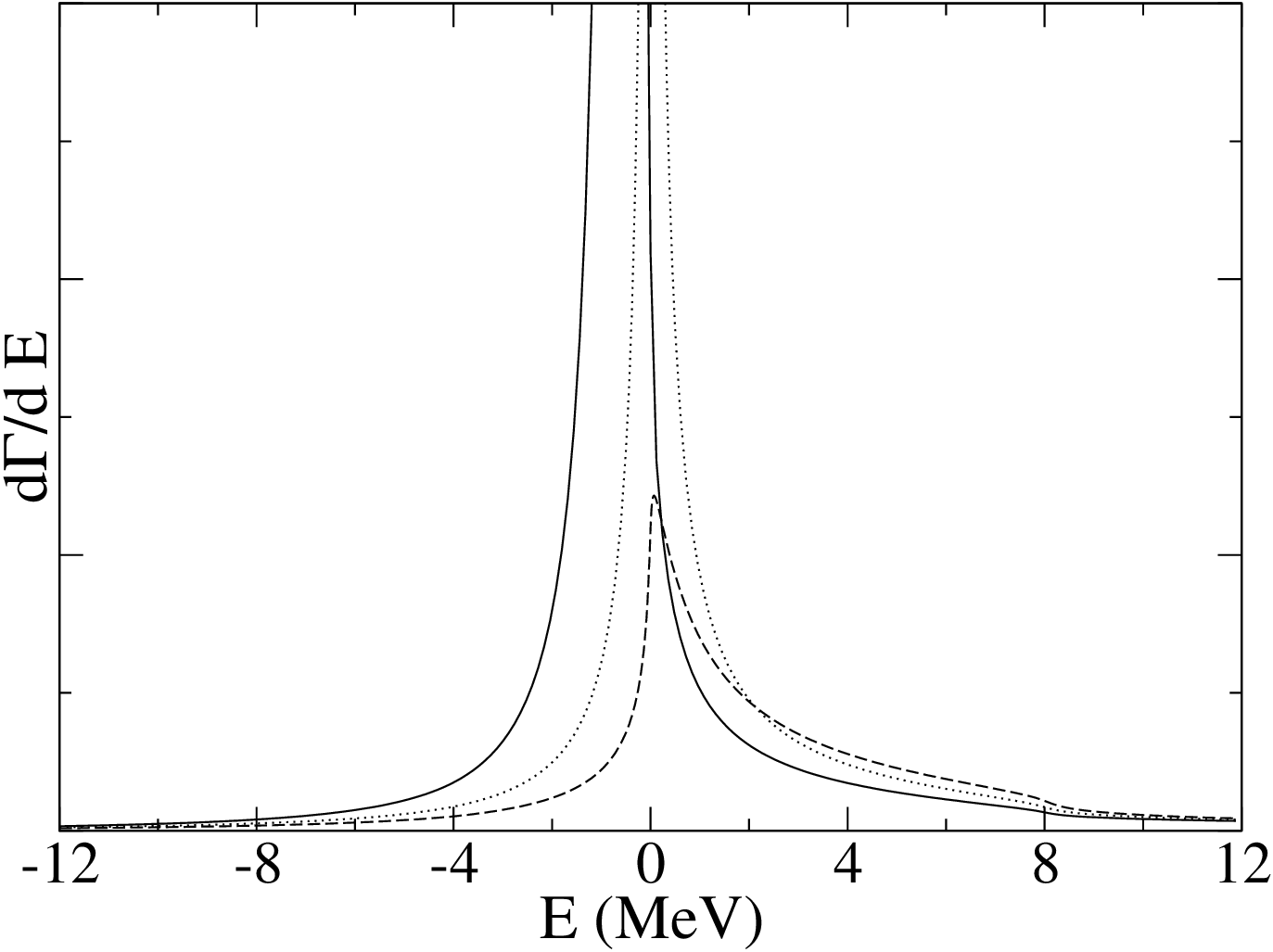}
\caption{The line shapes in the $D^{*} \bar D$ threshold region
for $X(3872)$ produced by a $B^+ \to K^+$ or $B^0 \to K^0$ transition 
and decaying into $J/\psi \, \pi^+ \pi^- \pi^0$.   
The line shapes are shown for $\gamma_1 = \pm \infty$
and three values of $\gamma$:
+34 MeV (solid line), 0 (dotted line), and $-34$ MeV (dashed line).
\label{fig:LS2BI0} }
\end{figure}

\begin{figure}[t]
\includegraphics[width=12cm,clip=true]{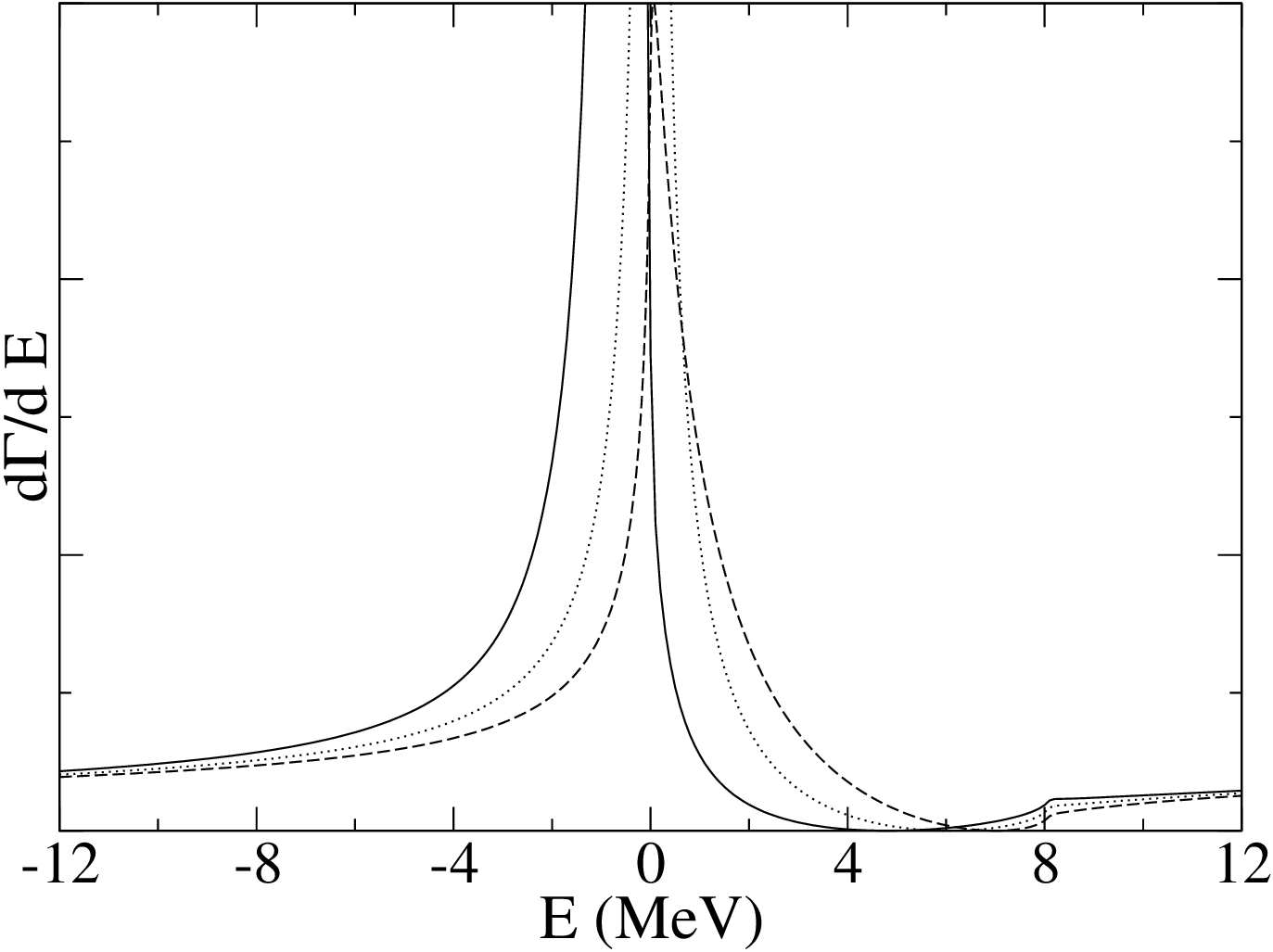}\\
\vspace{1cm}
\includegraphics[width=12cm,clip=true]{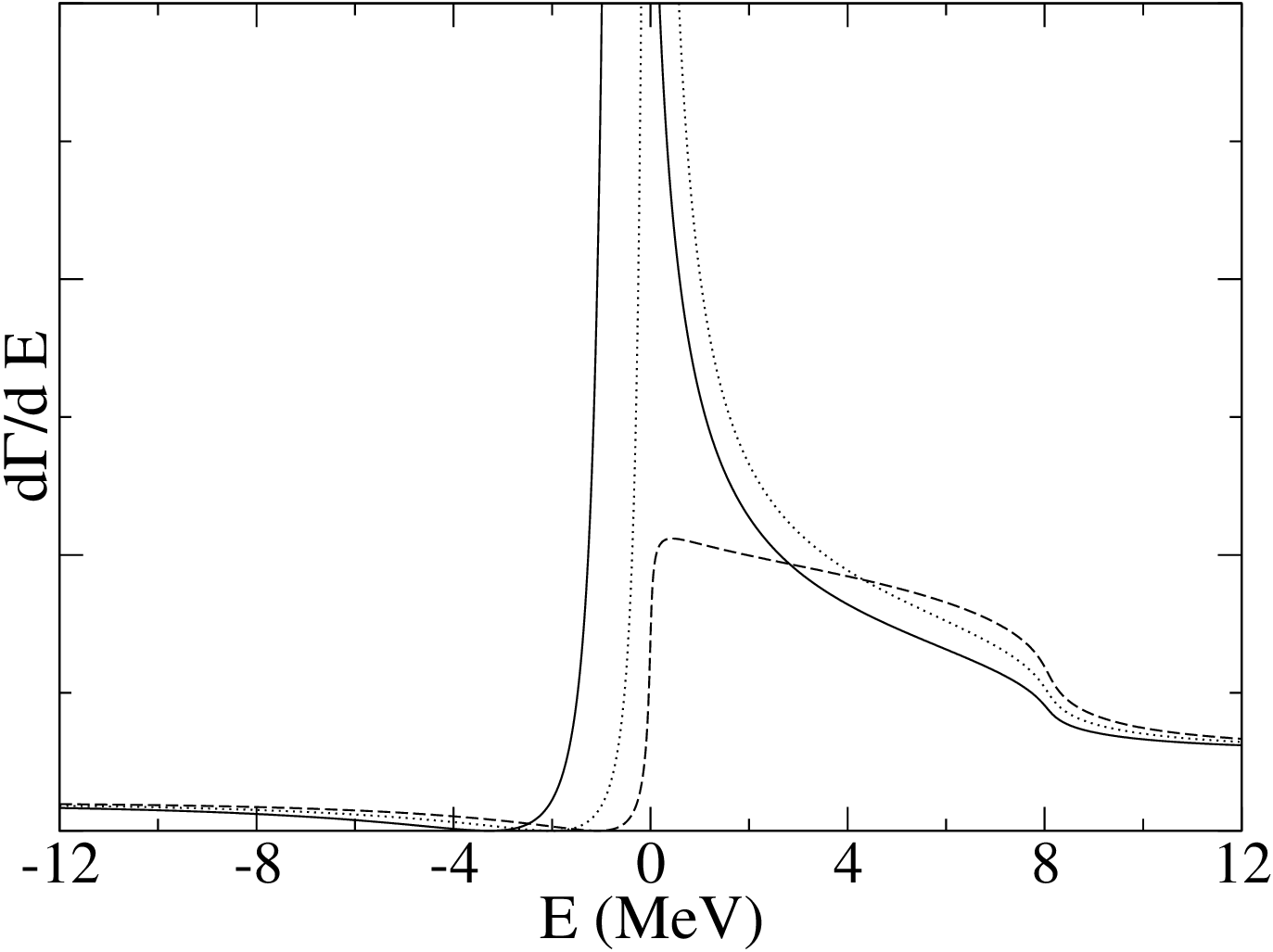}
\caption{The line shapes in the $D^{*} \bar D$ threshold region
for $X(3872)$ produced by a $B \to K$ transition 
and decaying into $J/\psi \, \pi^+ \pi^-$.  
The line shapes are different for $X$ produced by a 
$B^+ \to K^+$  transition (upper panel)
and a $B^0 \to K^0$ transition (lower panel).
The line shapes are shown for $\gamma_1 = \pm \infty$
and three values of $\gamma$:
+34 MeV (solid lines), 0 (dotted lines), and $-34$ MeV (dashed lines).
\label{fig:LS2B+0I1} }
\end{figure}

\begin{figure}[t]
\includegraphics[width=12cm,clip=true]{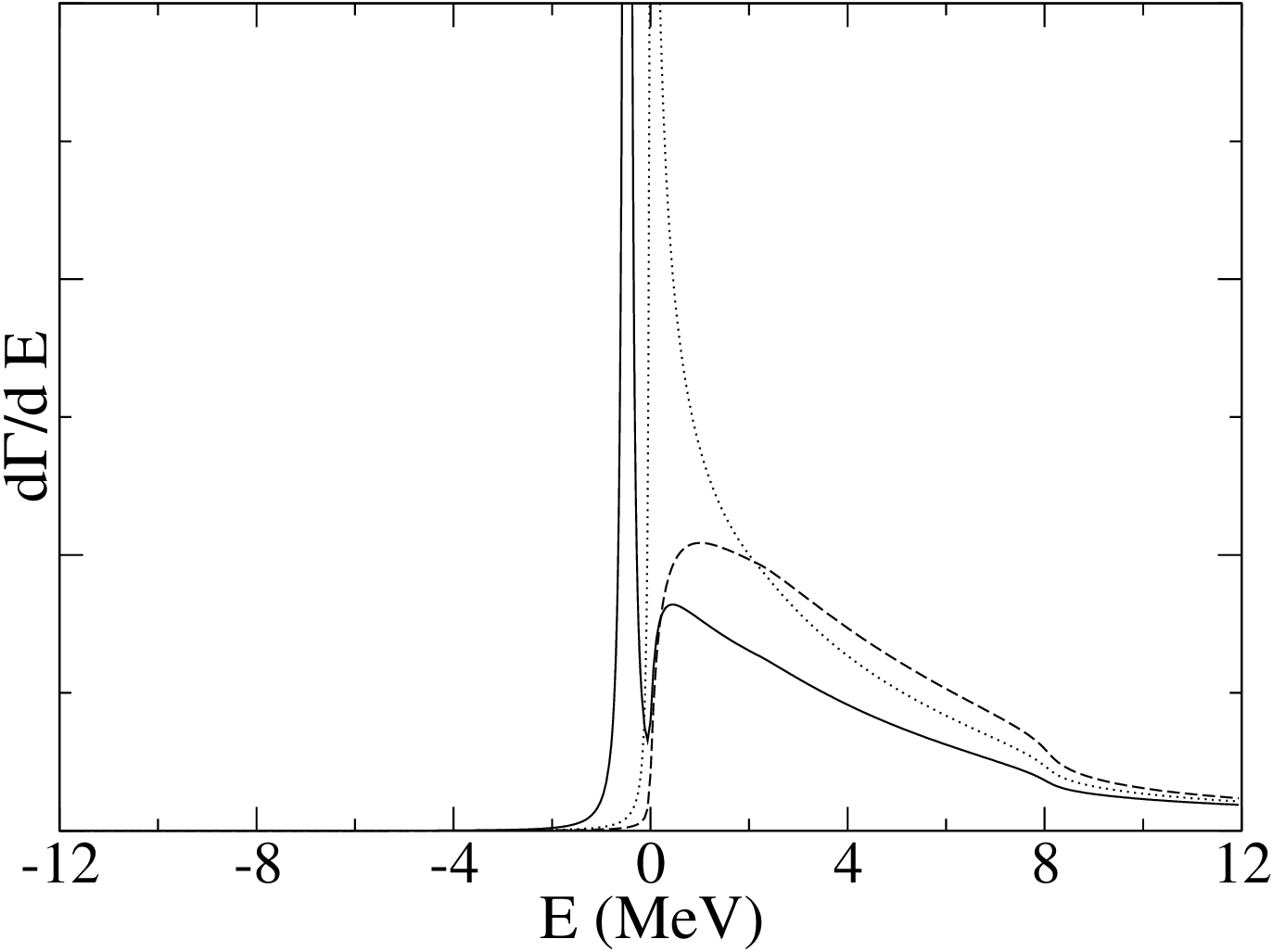}\\
\caption{The line shapes in the $D^{*} \bar D$ threshold region
for $X(3872)$ produced by a $B^+ \to K^+$ or $B^0 \to K^0$ transition 
and decaying into $D^0 \bar D^0 \pi^0$.  
The line shapes are shown for $\gamma_1 = \pm \infty$
and three values of $\gamma$:
+34 MeV (solid line), 0 (dotted line), and $-34$ MeV (dashed line).
\label{fig:LS2BDDpi} }
\end{figure}

In Figs.~\ref{fig:LS2BI0}, \ref{fig:LS2B+0I1}, and \ref{fig:LS2BDDpi},
we illustrate the line shapes in the $D^{*} \bar D$ threshold region
for $X(3872)$ produced by $B \to K$ transitions.
We take into account the $D^{*0}$ width, but we neglect the effect 
on the line shapes of inelastic scattering channels for the charm mesons.
For simplicity, we show only the line shapes for the limiting case
$\gamma_1 \to \pm \infty$.  Thus the denominators $D(E)$ can be 
approximated by Eq.~(\ref{D:largegam1}) and numerator factors 
such as $\gamma_1 - \kappa_1(E)$ or $\gamma_1 - \kappa(E)$ 
can be approximated by $\gamma_1$.
The parameters $\gamma$ and $\gamma_0$ are related by the pole equation 
$2 \gamma_0 - \gamma - \kappa_1(E_{\rm pole}) = 0$, where 
$E_{\rm pole}$ is given in Eq.~(\ref{Epole}).  If we take $\gamma _0$ 
to be real, then $\gamma$ has an unphysical negative imaginary part.
We therefore take $\gamma$ to be real and use the pole equation to
determine the complex parameter $\gamma_0$:
\begin{equation}
\gamma_0 = \frac12 
 \left( \sqrt{ 2 M_{*11} \nu + (M_{*11}/M_{*00}) \gamma^2 
 + i M_{*11} (\Gamma[D^{*0}] - \Gamma_{*1}(-\nu))} + \gamma \right) .
\end{equation}
We show the line shapes for three real values of $\gamma$:
+34, 0, and $-34$ MeV.  The corresponding values of $\gamma_0$ 
have real parts 82 MeV, 63 MeV, and 48 MeV, respectively.
Their imaginary parts are all approximately 0.00012 MeV, 
which is completely negligible.
For $\gamma = +34$ MeV, the peak of the resonance is at $E= -0.6$ MeV, 
which is the central value of the measurement in Eq.~(\ref{MX-CLEO}).

In Fig.~\ref{fig:LS2BI0}, we show the line shapes in the short-distance 
decay mode $J/\psi \, \pi^+ \pi^- \pi^0$.  The line shapes, 
which are the same for $X$ produced by a $B^+ \to K^+$ 
or $B^0 \to K^0$ transition, are given in Eqs.~(\ref{lsh2f:Bpsi3pi}).   
The relative normalizations of the curves for the three values 
of $\gamma$ are determined by using the same short-distance factors 
$\Gamma_{B^+}^{K^+,00}$ and $\Gamma^{J/\psi \, \pi^+ \pi^- \pi^0}$.  
In Fig.~\ref{fig:LS2B+0I1}, we show the line shapes in the short-distance 
decay mode $J/\psi \, \pi^+ \pi^-$.  The line shapes are given in
Eqs.~(\ref{lsh2f:Bpsi2pi}).  The upper and lower panels show the 
line shapes produced by $B^+ \to K^+$ and $B^0 \to K^0$ transitions,
respectively.  The line shapes from the $B^+ \to K^+$ transition
have approximate zeros near $+6$ MeV, while the line shapes from the 
$B^0 \to K^+$ transition have approximate zeros near $-2$ MeV.
The relative normalizations of all six curves are determined 
by using the same short-distance factors 
$\Gamma_{B^+}^{K^+,00}$ and $\Gamma^{J/\psi \, \pi^+ \pi^-}$. 
In Fig.~\ref{fig:LS2BDDpi}, we show the line shapes in 
$D^0 \bar D^0 \pi^0$.  The line shapes, which are the same for $X$ 
produced by a $B^+ \to K^+$ or $B^0 \to K^0$ transition, are given in
Eqs.~(\ref{lsh2f:BDDpi}).   
The relative normalizations of the curves for the three values 
of $\gamma$ are determined by using the same short-distance factor 
$\Gamma_{B^+}^{K^+,00}$.  

Fig.~\ref{fig:LS2B+0I1} illustrates the fact that the line shape
of the $X(3872)$ may depend not only on the decay channel 
but also on the production mechanism for the resonance.
The difference between the line shapes in the 
$J/\psi \, \pi^+ \pi^-$ decay channel for $X$ produced 
by $B^+ \to K^+$ and $B^0 \to K^0$ 
transitions is particularly dramatic because of the approximate 
zeros in the line shapes.  These approximate zeros are general 
features of the line shapes in Eqs.~(\ref{lsh2f:Bpsi2pi}).
If the imaginary parts of $\gamma_0$ and $\kappa_1(E)$
are neglected,
the numerator factor $|\gamma_0 - \kappa_1(E)|^2$ 
in the energy distribution in Eq.~(\ref{lsh2f:B+psi2pi})
has a zero between the $D^{*0} \bar D^0$ 
and $D^{*+} D^-$ thresholds. 
If $|\gamma| \ll |\kappa_1(0)| \ll |\gamma_1|$, the 
approximate expression for $\gamma_0$ in 
Eq.~(\ref{gamma0-approx}) reduces to $\kappa_1(0)/2$.  
The zero is therefore near $\frac 3 4 \, \nu \approx 6.1$ MeV.
If the imaginary parts of $\gamma_0$ and $\kappa(E)$
are neglected,
the numerator factor $|\gamma_0 - \kappa(E)|^2$ 
in the energy distribution in Eq.~(\ref{lsh2f:B0psi2pi})
has a zero below the $D^{*0} \bar D^0$ threshold. 
If $|\gamma| \ll |\kappa_1(0)| \ll |\gamma_1|$, 
the zero is near $- \frac 1 4 \, \nu \approx -2.0$ MeV.
In the case of $B^+$ decays, the approximate zero forces 
the line shape to be narrower on the trailing edge of the resonance.
In the case of $B^0$ decays, the approximate zero forces 
the line shape to be narrower on the leading edge of the resonance.

In Ref.~\cite{Braaten:2004ai}, Braaten and Kusunoki predicted that 
the decay rate for $B^0 \to K^0 + X(3872)$ should be suppressed 
relative to that for $B^+ \to K^+ + X(3872)$.  
Their prediction was based on the current-current factorization 
approximation and heavy-quark symmetry.  Together they
imply that, in the $D^*\bar D$ threshold region,
the $B^+ \to K^+$ transition creates charm mesons predominantly 
in the neutral channel $(D^* \bar D)_+^0$, while the
$B^0 \to K^0$ transition creates them predominantly
in the charged channel $(D^* \bar D)_+^1$.  Since the $X(3872)$ is a 
resonance in the $(D^* \bar D)_+^0$ channel, the authors
of Ref.~\cite{Braaten:2004ai} concluded that the rate for 
$B^0 \to K^0 + X$ must be suppressed relative to that for 
$B^+ \to K^+ + X$.  In retrospect,  this prediction was the result of
a conceptual error. 

The conclusion of Ref.~\cite{Braaten:2004ai} that $B^0 \to K^0 + X$ 
is suppressed follows from the factorization formulas in 
Sections~\ref{sec:lineshape2}
if $\kappa_1(0)$ is assumed to be much greater than $\gamma_0$ and
$\gamma_1$. In this limit, the two-channel factorization formulas 
for $B^+$ decays in Eqs.~(\ref{lsh2f:B+psi3pi}), (\ref{lsh2f:B+psi2pi}), 
and (\ref{lsh2f:B+DDpi}) all reduce to the single-channel factorization 
formulas in Eqs.~(\ref{lsh1:C}) and (\ref{lsh1:DDbarpi}), where $f(E)$ 
is the single-channel scattering amplitude in Eq.~(\ref{f-E}), 
$\gamma = (\gamma_1 + \gamma_0)/2$, $\Gamma (E) = 2 \Gamma_I (E)$, 
and $\Gamma^{K^+}_{B^+} = \Gamma^{K^+, 00}_{B^+}$.  For the inverse
scattering length $\gamma$ to be small compared to $\gamma_0$ and
$\gamma_1$, $\gamma_0$ and $\gamma_1$ must be nearly equal in magnitude
but opposite in sign. If $\gamma$ is small
compared to $\gamma_1$, the factorization formula for $B^0$ decays in
Eqs.~(\ref{lsh2f:B0psi3pi}), (\ref{lsh2f:B0psi2pi}), and
(\ref{lsh2f:B0DDpi}) also reduce to the single-channel
factorization formulas in Eqs.~(\ref{lsh1:C}) and (\ref{lsh1:DDbarpi}),
with $\Gamma^{K^0}_{B^0} = [\gamma_1/\kappa_1(0)]^2 \Gamma^{K^+}_{B^+}$. 
The rate for $B^0 \to K^0 + X$ is therefore suppressed by a 
factor of $[\gamma_1 /\kappa_1(0)]^2$ compared to the rate for 
$B^+ \to K^+ +X$. Thus the
conclusion of Ref.~\cite{Braaten:2004ai} is consistent with the factorization
formulas only if the parameters satisfy the hierarchy $|\gamma| \ll
|\gamma_1| \ll |\kappa_1 (0)|$. Since the conclusion of 
Ref.~\cite{Braaten:2004ai} is only valid in one corner of the parameter
space, the authors must have made a conceptual error.

The conceptual error in Ref.~\cite{Braaten:2004ai}
has to do with the momentum scale at which the inferences from the 
current-current factorization approximation and heavy-quark symmetry
are applied. Heavy quark symmetry
is relevant at energy scales that are small compared to 
the heavy quark mass $m_c$ and large compared to the energy scale 
set by isospin symmetry violations, which is $\nu \approx 8.1$ MeV.
Equivalently, it is relevant at momentum scales that are small 
compared to $m_c$ and large compared to $\kappa_1(0) \approx 125$ MeV.
The current-current factorization approximation and heavy quark symmetry
imply that the short-distance constant
$\Gamma^{K^+, 00}_{B^+}=\Gamma^{K^0, 11}_{B^0}$ dominates. 
This inference should be applied at a
momentum scale where heavy quark symmetry applies, which requires the
momentum to be large compared to $\kappa_1 (0) \approx 125$ MeV. The
conceptual error in Ref.~\cite{Braaten:2004ai} was inferring that 
the transitions $B^+ \to K^+ + (D^* \bar D)^0_+$ and 
$B^0 \to K^0 + (D^* \bar D)^1_+$ 
dominate at a momentum scale small compared to $\kappa_1 (0)$. 
At this low momentum scale, there is a
resonance only in the $(D^* \bar D)^0_+$ channel. However the dominance
of $\Gamma^{K^+, 00}_{B^+}=\Gamma^{K^0, 11}_{B^0}$ at a scale large
compared to $\kappa_1 (0)$ does not imply the dominance of
the transitions $B^+ \to K^+ + (D^* \bar D)^0_+$ and 
$B^0 \to K^0 + (D^* \bar D)^1_+$ at lower
scales. As the momentum scale is lowered, resonant scattering between the
$(D^* \bar D)^0_+$ and $(D^* \bar D)^1_+$ channels can feed the
transitions $B^+ \to K^+ + (D^* \bar D)^1_+$ and 
$B^0 \to K^0 + (D^* \bar D)^0_+$.

To deduce the correct implications of the current-current factorization
approximation and heavy quark symmetry at momentum scales small compared
to $\kappa_1 (0)$, we can consider the general factorization formulas in
Eqs.~(\ref{lsh2f:Bpsi3pi}), (\ref{lsh2f:Bpsi2pi}), and
(\ref{lsh2f:BDDpi}) in the low energy region where $\kappa (E)$
is small compared to $\kappa_1(0)$. In this region, the factorization
formulas reduce to the single-channel factorization formulas in
Eqs.~(\ref{lsh1:C}) and (\ref{lsh1:DDbarpi}). 
The assumption that $\Gamma^{K^+, 00}_{B^+} = \Gamma^{K^0, 11}_{B^0}$ 
dominates implies that the short-distance
constants for the $B \to K$ transition are 
$\Gamma^{K^+}_{B^+} \approx \Gamma^{K^+, 00}_{B^+}$ and 
$\Gamma^{K^0}_{B^0} \approx \Gamma^{K^+, 00}_{B^+} \, |c_1|^2$, where
the coefficient $c_1$ is given in Eq.~(\ref{c1-factor}). 
The ratio of the rates is 
$|c_1|^2 = |\gamma_1|^2 / | \gamma_1 - \kappa_1(0)|^2$.
If we ignore the small imaginary parts of $\gamma_1$ and $\kappa_1(0)$, 
this ratio is greater than 1 if $\gamma_1 > \kappa_1(0)/2$ and less than 1
if $\gamma_1 < \kappa_1(0)/2$.
Thus the rate for $B^0 \to K^0 + X$ need not be suppressed 
compared to that for $B^+ \to K^+ +X$.

\section{Summary}
\label{sec:summary}

In Ref.~\cite{Braaten:2007dw}, we derived line shapes of the $X(3872)$ 
that should be accurate in the
region within a few MeV of the $D^{*0} \bar D^0$ threshold.
The line shapes were derived from an expression for the
resonant scattering amplitude in the $(D^* \bar D)^0_+$ channel
that takes into account the $D^{*0}$ width and inelastic charm
meson scattering channels. In the factorization formulas for the line
shapes, short-distance effects and long-distance effects are separated
into multiplicative factors. The line shapes of Ref.~\cite{Braaten:2007dw} 
are independent of the production mechanism for the $X$ resonance.
The line shape in $D^0 \bar D^0 \pi^0$ is different from the line shape 
in a short-distance decay mode, such as $J/\psi \, \pi^+ \pi^- \pi^0$ 
or $J/ \psi \, \pi^+ \pi^-$. As
shown by the analysis of Ref.~\cite{Braaten:2007dw},
the difference in these line shape
can explain the difference between the masses of $X(3872)$ measured
in the $J/ \psi \, \pi^+ \pi^-$ and $D^0 \bar D^0 \pi^0$ decay modes
\cite{Gokhroo:2006bt,Babar:2007rv}.

In this paper, we have derived line shapes for the $X(3872)$ whose region
of validity extends to the entire $D^* \bar D$ threshold region by taking
into account the resonant coupling between the $(D^* \bar D)^0_+$ and
$(D^* \bar D)^1_+$ channels. By taking into account isospin symmetry at
high energies, the coupled-channel scattering amplitudes were expressed
in terms of two parameters: the $I = 0$ and $I = 1$ inverse scattering
amplitudes $\gamma_0$ and $\gamma_1$. Isospin symmetry was also taken
into account in the short-distance factors in the factorization formulas.
In the case of production of the $X$ resonance in $B \to K$ transitions,
isospin symmetry reduces the short-distance factors to four independent real
constants: $\Gamma_{B^+}^{K^+,00}$, $\Gamma_{B^+}^{K^+,11}$, and the
real and imaginary parts of $\Gamma_{B^+}^{K^+,01}$. The resulting
factorization formulas for the inclusive resonance production rate in
the $B^+ \to K^+$ transition is given in Eq.~(\ref{lsh2:res}).
The factorization formula for the $D^0 \bar D^0 \pi^0$ channel is given
in Eq.~(\ref{lsh2:DDbarpi}). The factorization formulas for $I=0$ and $I=1$
short-distance decay channels are given for both the $B^+ \to K^+$ 
and the $B^0 \to K^0$ transitions in Eqs.~(\ref{lsh:BI=0}) 
and (\ref{lsh:BI=1}).  The line shape in an $I=0$ short-distance decay channel, 
such as $J/\psi \, \pi^+ \pi^- \pi^0$, is different from the line shape 
in an $I=1$ short-distance decay channel, such as $J/ \psi \, \pi^+ \pi^-$.
The line shapes for the $X$ resonance produced by the $B^+ \to K^+$ 
transition are also different from the line shapes produced by the 
$B^0 \to K^0$ transition. 

If we use the current-current factorization approximation together with
heavy quark symmetry, the factorization formulas simplify dramatically.
The factorization formulas for an $I = 0$ decay channel, an $I=1$ decay
channel, and $D^0 \bar D^0 \pi^0$ are given in Eqs.~(\ref{lsh2f:Bpsi3pi}),
(\ref{lsh2f:Bpsi2pi}), and (\ref{lsh2f:BDDpi}), respectively. 
The short-distance constants
associated with the $B\to K$ transitions reduce to a single real constant
$\Gamma_{B^+}^{K^+,00}$. Thus the ratios of production rate in 
$B^+ \to K^+$ transitions and in $B^0 \to K^0$ transitions 
are completely determined by the
scattering parameters $\gamma_0$ and $\gamma_1$.

Our results allow us to identify conceptual errors in previous work on
this problem. In Ref.~\cite{Braaten:2004ai}, 
Braaten and Kusunoki predicted that $B^0
\to K^0 + X$ should be suppressed by at least an order of magnitude
compared to $B^+ \to K^+ +X$. The prediction was based on the
current-current approximation and heavy quark symmetry. The conceptual
error was an implicit assumption that $\gamma_0$ and $\gamma_1$ are small
compared to $\kappa_1 (0)$. In Ref.~\cite{Voloshin:2007hh}, 
Voloshin predicted that the
lines shapes for the $X$ resonance produced by $B^+ \to K^+ + X$ and 
$B^0 \to K^0 + X$ should be identical. The conceptual errors were ignoring
resonant scattering between the $(D^* \bar D)^0_+$ and $(D^* \bar D)^1_+$
channels and failing to take into account isospin symmetry in the 
$B \to K$ transitions.

Our results provide a physical interpretation for the model of the 
$(D^* \bar D)^0_+$ scattering amplitude used in Ref.~\cite{Hanhart:2007yq}. 
The scaling behavior of the fits in Ref.~\cite{Hanhart:2007yq} indicate that 
the term $- (2/g)E$ in the inverse of the scattering amplitude 
in Eq.~(\ref{f:HKKN}) can
be omitted. The resulting scattering amplitude is essentially equivalent
to the general $(D^* \bar D)^0_+$ scattering amplitude $f_{00}(E)$ in
Eq.~(\ref{f00-E}) in the limit $|\gamma_1 | \gg |\kappa_1 (0)|$.

The most important parameters for predicting the line shapes are the
scattering parameters $\gamma_0$ and $\gamma_1$. They could be determined
phenomenologically from ratios of rates for $B^0 \to K^0 +X$ and 
$B^+ \to K^+ + X$. Alternatively they could be calculated using the meson
potential model of Ref.~\cite{Tornqvist:1993ng}. If these scattering parameters 
were calculated, the predictive
power of the results of this paper would be dramatically increased.
The line shapes for the $X(3872)$ resonance produced by $B \to K$ transitions
also depend on the short-distance factors $\Gamma_{B^+}^{K^+,00}$,
$\Gamma_{B^+}^{K^+,01}$, and $\Gamma_{B^+}^{K^+,11}$.
They could be determined phenomenologically from measurements of the 
charm meson invariant mass distributions in the decays 
$B \to K + D^* \bar D$ and $B \to K + D \bar D^*$. 

The accuracy of our predictions for the line shapes could be 
further improved by taking into account pions explicitly.
The system consisting of $D^{*0} \bar D^0$,
$D^0 \bar D^{*0}$, and $D^0 \bar D^0 \pi^0$ states with energies
near the $D^0 \bar D^{*0}$ threshold can be described by a 
nonrelativistic effective field theory.  The simplest such theory 
has S-wave scattering in the $(D^* \bar D)_+^0$ channel
and $\pi^0$ couplings that allow the decay $D^{*0} \to D^0 \pi^0$.
Fleming, Kusunoki, Mehen, and van Kolck developed power-counting 
rules for this effective field theory and showed that the pion
couplings can be treated perturbatively \cite{Fleming:2007rp}.
They used the effective field theory to calculate the decay rate for
$X(3872) \to D^0 \bar D^0 \pi^0$ to next-to-leading order in the 
pion coupling.
In applying this effective field theory to the line shapes of the 
$X(3872)$, one complication that will be encountered is infrared 
singularities at the $D^{*0} \bar D^0$ threshold that are related 
to the decay $D^{*0} \to D^0 \pi^0$.  This problem has been 
analyzed in a simpler model with spin-0 particles and 
momentum-independent interactions \cite{Braaten:2007ct}.
The problem was solved by a resummation of perturbation theory
that takes into account the perturbative shift of the 
$D^0 \bar D^{*0}$ threshold into the complex energy plane
because of the nonzero width of the $D^{*0}$.

In summary, the establishment of the quantum numbers of the 
$X(3872)$ as $1^{++}$ and the measurement of its mass imply 
that it is either a charm meson molecule or a charm meson 
virtual state.  These two possibilities can be distinguished 
in practice by their different predictions for the line shapes 
of the $X(3872)$.  The analysis of Ref.~\cite{Braaten:2007dw} 
indicates that the existing data favor a charm meson molecule,
but a virtual state is not excluded.
The expressions for the line shapes used in that analysis
should be accurate only
within a few MeV of the $D^{*0} \bar D^0$ threshold.
The expressions for the line shapes derived in this paper 
should be accurate in the entire $D^* \bar D$ threshold region.
When more extensive data on the line 
shapes of the $X(3872)$ in various decay channels 
and for various production processes becomes available,
it should be possible to determine conclusively whether the $X(3872)$ 
is a bound state or a virtual state of charm mesons.

\begin{acknowledgments}
This research was supported in part by the Department of Energy
under grant DE-FG02-91-ER40690.
\end{acknowledgments}



\end{document}